\documentclass[10pt,aps,prd,nofootinbib,superscriptaddress,twocolumn,preprintnumbers,balancelastpage]{revtex4-2}

\usepackage[colorlinks=true
,urlcolor=blue
,anchorcolor=blue
,citecolor=blue
,filecolor=blue
,linkcolor=blue
,menucolor=blue
,linktocpage=true
,pdfproducer=medialab
,pdfa=true
]{hyperref}
\usepackage{amssymb,amsmath,multirow,slashed,cleveref} 

\usepackage{graphicx}
\usepackage{lipsum}
\usepackage{bm}
\usepackage{setspace}
\usepackage[T1]{fontenc}

\newcommand{\ket}[1]{\ensuremath{| #1 \rangle }}

\begin{document}
\preprint{CERN-TH-2024-104}

\title{A Nuclear Interferometer for Ultra-Light Dark Matter Detection }

\author{Hannah Banks}
\affiliation{CERN, Theoretical Physics Department, Geneva, Switzerland}
\affiliation{DAMTP, University of Cambridge, Wilberforce Road, Cambridge, UK}
\author{Elina Fuchs}
\affiliation{CERN, Theoretical Physics Department, Geneva, Switzerland}
\affiliation{Institute of Theoretical Physics, Leibniz Universität Hannover, Appelstr.~2, 30167 Hannover, Germany}
\affiliation{Physikalisch-Technische Bundesanstalt (PTB), Bundesallee 100, 38116 Braunschweig, Germany}
\affiliation{Deutsches Elektronen-Synchrotron DESY, Notkestr. 85, 22607 Hamburg, Germany}
\author{Matthew McCullough}
\affiliation{CERN, Theoretical Physics Department, Geneva, Switzerland}

\begin{abstract}
\begin{center}
{\bf Abstract}
\vspace{-7pt}
\end{center}
We propose the nuclear interferometer - a single-photon interferometry experiment based upon the thorium-229 nuclear clock transition - as a novel detector for ultra-light dark matter. Thanks to the enhanced sensitivity of this transition to the variation of fundamental constants, we find that possible realisations of such an experiment deploying either single ions or clouds of atoms have the potential to complement advanced very-long-baseline terrestrial clock atom interferometers in the search for  ultra-light dark matter with scalar couplings to photons in the future. Nuclear interferometry may also offer an unparalleled window to new physics coupling to the QCD sector via quarks or gluons, with a discovery reach that could enhance existing and proposed experiments over a range of frequencies in the direction of well-motivated parameter space.

\end{abstract}

\maketitle


\section{Introduction}
The hunt for dark matter (DM) is entering a new era. Despite a panoply of evidence for its existence, very little is known about the nature or properties of DM on the particle level. At present, our only real handle on this elusive substance comes from observations of its gravitational interactions over astrophysical and cosmological scales, leaving its microphysics largely unconstrained. Even under the assumption that the energy density attributed to DM is predominantly harboured by a single species, the mass in question could viably range from   $\sim 10^{-22}$\,eV bosons up to $\sim 10^{19}$\,GeV mass primordial black holes~\cite{Battaglieri:2017aum,Green:2020jor}. Deciphering the fundamental constituents of DM and its particle-like properties remains of upmost priority in contemporary particle physics. 

 Over the last few decades, non-gravitational interactions of weak-scale particles have been relentlessly pursued as part of a broad experimental campaign encompassing direct, indirect and collider based searches, all to no avail~\cite{ParticleDataGroup:2022pth}. Whilst there remain a number of interesting and well-motivated avenues in the vicinity of this parameter space yet to explore, the idea that DM instead comprises sub-eV ultra-light bosonic particles which have remained hidden thus far due to extremely feeble couplings to the Standard Model (SM), has gained significant traction in recent years~\cite{Jaeckel:2010ni,Antypas:2022asj}. A number of highly motivated DM candidates fall into this category including the QCD axion~\cite{Peccei:1977hh,Shifman:1979if} (see Ref.~\cite{DiLuzio:2020wdo} for a review) and other pseudo-scalar axion-like particles (ALPs), vector dark photons~\cite{Holdom:1985ag,Fayet:1990wx,Fabbrichesi:2020wbt} and scalars such as moduli~\cite{Dimopoulos:1996kp,Arkani-Hamed:1999lsd,Burgess:2010sy,Cicoli:2011yy}, dilatons~\cite{Damour:1994zq,Taylor:1988nw}, Higgs portal dark matter~\cite{Piazza:2010ye} and the relaxion~\cite{Graham:2015cka,Banerjee:2018xmn}. 
 
 The high occupation numbers required for such a light species to saturate the relic local dark matter density renders its behaviour over galactic scales akin to that of a coherent, classical wave,  giving rise to a host of novel phenomenological signatures which are not shared by heavier `particle-like' candidates~\cite{Ferreira:2020fam,Hui:2021tkt}.  Direct searches for ultra-light DM (ULDM) thus typically rely upon different detection strategies to the traditional experiments conceived for weak-scale candidates which seek impulses resulting from particle-like interactions.  Recent advances in quantum sensing technologies have been transformative in this respect, opening up a number of exciting new possibilities to probe new light weakly coupled states which lie within this `precision' frontier~\cite{qt_elephants,Safronova:2017xyt,Buchmueller:2022djy,Chou:2023hcc,Bass:2023hoi}. 
 
 ULDM could conceivably couple to the SM through a number of different `portals', each of which has a distinct set of phenomenological consequences and in turn, experimental signatures. In this proposal we focus specifically on \textit{scalar} ULDM with linear couplings to the SM. These couplings cause fundamental constants, and thus physical properties such as atomic energy levels and the distance scales which they govern, to oscillate in time at a frequency set by the DM mass, see e.g. Refs.~\cite{Uzan:2010pm,Arvanitaki:2014faa,Stadnik:2014tta,Stadnik:2015kia,Safronova:2017xyt}. This effect can be searched for in a variety of different ways including accelerometers \cite{PhysRevD.93.075029}, optical cavities~\cite{PhysRevLett.123.031304}, mechanical resonators~\cite{Manley:2019vxy,Carney:2020xol} and  by comparing the frequency of two different atomic (see e.g. Refs.~\cite{Arvanitaki:2014faa,Stadnik:2015kia,BACON:2020ubh,Filzinger:2023zrs}) or molecular~\cite{Madge:2024aot} clocks. More recently, atom interferometers have been identified to offer the potential to probe ULDM-induced oscillations of  atomic transition frequencies in a complementary region of parameter space to the aforementioned tests~\cite{Geraci:2016fva,Arvanitaki:2016fyj}.  In its most basic form, an atom interferometer constitutes an experiment which measures the phase difference between two spatially delocalised quantum superpositions of clouds of atoms \cite{Abend:2020djo}. Such experiments can be thought of analogously to optical interferometers but with the  traditional beam-splitters and mirror elements realised via a sequence of controlled duration laser pulses which act to divide, reflect and recombine the atomic wave-packets by driving transitions between internal energy states. The presence of ULDM can be diagnosed by the existence of an oscillating  phase difference between a pair of spatially separated single-photon\footnote{Here, `single-photon' refers specifically to a class of interferometers in which the desired transition is driven by a single laser frequency. This contrasts with interferometers based on Raman transitions which involve two distinct optical fields (see e.g. Ref.~\cite{PhysRevD.78.122002}).} atom interferometers which are resonantly interrogated with common laser pulses \cite{Arvanitaki:2016fyj}.  Such multi-atom interferometer configurations are termed single-photon atom gradiometers and have the key advantage in that noise arising from fluctuations and jitters in the laser phase or mechanical vibrations is common to both instruments and thus cancels in the differential measurement \cite{PhysRevLett.110.171102}. In order to achieve sensitivity to ULDM, such proposals look to both long ($\mathcal{O}$(km)) baselines and large-momentum-transfer (LMT) atom optics\footnote{Such techniques (see e.g. Refs.~\cite{Mazzoni:2015kpa,Rudolph:2019vcv,Kirsten-Siemss:2022kzy}) effect an increase of the physical baseline by employing an increased number of laser pulses to augment the momentum transfer between the two arms of each atom interferometer and in turn the phase difference accumulated at the end of the sequence.} \cite{PhysRevLett.85.4498}  (see Sec.~\ref{sec:sig}) to increase the recorded phase difference.  To sustain these practices it is essential for the pair of internal states on which the atom interferometer is based to be sufficiently long-lived such that spontaneous emission, which would otherwise exponentially degrade the sensitivity due to loss of coherence, is not significant.\footnote{For a discussion of some of the potential implications of spontaneous decay on long-baseline atom interferometry in space, see Refs.~\cite{PhysRevD.89.062004,Loriani:2018qej}.} For the very long baselines and high number of sequential pulses demanded by ULDM searches (see e.g.\ \cite{Graham:2017pmn,Badurina:2021rgt}), ultra-narrow transitions such as those deployed in optical lattice clocks are therefore typically taken advantage of. For this reason such atom interferometer realisations have earned the title of `large-momentum-transfer clock atom interferometry' \cite{Proceedings:2023mkp}.
 
 A number of prototype long-baseline gradiometers including AION-10 \cite{Badurina:2019hst}, MAGIS-100 \cite{Abe_2021} and VLBAI~\cite{Hartwig_2015} are currently in development, and will serve as essential technological readiness indicators to longer (up to km-scale) instruments envisioned to offer sensitivity to oscillating ULDM as well as mid-frequency (0.1-10 Hz) gravitational waves by the mid 2030's. A comprehensive overview of the potential capabilities of long-baseline interferometers in addition to their current status of development can be found in Ref.~\cite{Proceedings:2023mkp}.
 
In tandem, the discovery of an excited isomeric state in $^{229}$Th ~\cite{vonderWense:2016fto,Thirolf:2024_50yearsTh229m}, has led to proposals for a  `nuclear clock'~\cite{Peik:2003_ThProposal,PhysRevLett.108.120802,PhysRevLett.104.200802} based on transitions between the ground and the $^{229m}$Th excited isomer state. Arising due to accidental cancellations between contributions from  strong and electromagnetic interactions \cite{PhysRevLett.97.092502,Flambaum:2008ij,PhysRevLett.102.210801}, the small (8.3 eV \cite{Seiferle:2019fbe}) gap in energy between the ground and excited states  lies in the optical range~\cite{Beck:2007zza} and offers an enhanced sensitivity to the variation of fundamental constants at the level of several orders of magnitude\footnote{Note that the estimated degree of enhancement depends on the way that the nucleus is modeled and could, in the `nightmare' scenario be  $\mathcal{O}(1)$  (i.e. no enhancement) \cite{Caputo:2024doz}. As argued in this reference however, this situation is thought to be unlikely with enhancements of $\sim 10^4$ deemed most probable.} \cite{PhysRevLett.97.092502,Berengut_2011,Fadeev:2020bgt,Caputo:2024doz} over typical optical clock transitions in atoms, ions and highly charged ions~\cite{Kozlov:2018mbp}. Frequency comparisons between nuclear and atomic optical clocks could offer a highly sensitive probe of ULDM in the near future \cite{Peik:2020cwm,Thirolf:2019knx,Derevianko:2013oaa,Arvanitaki:2014faa,Banerjee:2020kww}, and the (still broad) lineshape already constrains the ULDM parameter space~\cite{Fuchs:2024edo}. Due to rapid  internal conversion, the excited isomeric state in electrically neutral $^{229}$Th is short-lived with a half-life of $\sim 7\,\mu$s \cite{Seiferle:2017lya}. Proposed realisations of the thorium nuclear clock for deployment in such experiments therefore look to either ionic forms \cite{PhysRevLett.108.120802} or to a thorium-doped lattice \cite{Kazakov_2012} in which internal conversion is forbidden and the isomeric state becomes  long-lived with a radiative lifetime of $\sim 10^4$ s \cite{Peik:2020cwm}.


Whilst the nuclear clock is yet to be realised experimentally, rapid progress towards this goal has been made in recent years and especially months. 
The production of $^{229}$Th at ISOLDE at CERN via the decay chain of $^{229}$Fr as a new pathway~\cite{Kraemer:2022gpi} refined the knowledge of the nuclear transition energy from the percent to the per-mill level. 
Subsequently, based on the development of a new tunable VUV laser~\cite{Thielking2023_VUVlaser_NJP}, the first laser excitation of the $^{229}$Th nucleus was achieved at PTB~\cite{Tiedau:2024obk} in Th-doped crystals~\cite{Beeks:2022dnl}. While the resolution of the transition frequency here is limited by the laser linewidth to 7\,GHz~\cite{Tiedau:2024obk}, corresponding to a relative uncertainty of $\mathcal{O}(10^{-6})$, this precision has since further been improved to 3\,GHz at UCLA~\cite{Elwell:2024qyh} and, by means of a VUV frequency comb at JILA, to 300\,kHz~\cite{Zhang:2024ngu},  suppressing the relative uncertainty to $\mathcal{O}(10^{-10})$. 
At a temperature where the temperature shift vanishes, Ref.~\cite{Ooi:2025hmd} reports a frequency stability of 280\,Hz (which corresponds to a fractional frequency stability of $10^{-13}$) comparing 2 different crystals over 4 months and finds a linear dependence of the linewidth on the doping concentration of the crystal. Co-thermometry is expected to reduce the fractional uncertainty from the temperature shift below $10^{-18}$.
In parallel, nuclear spectroscopy in Thorium ions of different charge provides promising prospects~\cite{Flambaum:2025xqs,Koziol:2024eaw}.
With such developments rapidly paving the way towards an operational nuclear clock, the time to investigate other ways in which this unique transition could be exploited to probe fundamental physics is ripe. 

In this work we propose to amalgamate the $^{229}$Th  nuclear clock transition with the established principles of  single-photon matter-wave interferometry to form what we shall refer to as a `nuclear interferometer'. We present two different possible realisations of this idea - one based on  single  $^{229}$Th ions and the other on clouds of neutral  $^{229}$Th atoms. Whilst both present additional experimental challenges beyond those of the optical clock transitions of conventional atom interferometry-based  ULDM search proposals, we find that for certain experimental configurations, the enhanced sensitivity of the ground-to-excited state transition to the variation of fundamental constants in these thorium  systems is able to overcome the loss in sensitivity that respectively results from the low ion flux (i.e large quantum projection noise) and the short  lifetime of the excited state in neutral form, to offer access to  new regions of  ULDM parameter space.

We begin in Sec.~\ref{sec:model} by introducing the ULDM model that we will adopt as an example in this work, before computing the signal that it would imprint in a generic nuclear interferometry experiment in Sec.~\ref{sec:sig}. In Sec.~\ref{sec:sens} we then derive an expression for the reach of a general free-fall nuclear interferometer assuming shot-noise limited detection, in terms of a handful of tunable parameters which characterise the experimental operating configurations. In Sec.~\ref{sec:real} we explore two different ways, making use of single-ion clocks and clouds of atoms respectively, in which  such a nuclear interferometer could be realised experimentally and discuss how the aforementioned principles apply to each case. The challenges of working with both systems are discussed and any requisite technological advancements outlined.   Forecasts of the sensitivity of both realisations to scalar ULDM with linear couplings to SM operators for a variety of experimental configurations that may become feasible in the future are presented in Sec.~\ref{sec:diss} and  compared to  bounds from existing experiments and the predicted sensitivities of competing proposals. Some concluding remarks are offered in Sec.~\ref{sec:sum}.

\label{sec:intro}

\section{The Concept}
\subsection{Ultra-light scalar Dark Matter}\label{sec:model}
To motivate this proposal we consider an ultra-light singlet scalar field, $\phi$, which  couples linearly\footnote{Note that quadratic  couplings also lead to the variation of fundamental constants \cite{Jiang:2024agx,Banerjee:2022sqg}, becoming relevant if linear couplings are not present, forbidden for instance by a symmetry. We consider the case of the QCD axion, whose defining anomalous coupling to gluons generates an oscillatory contribution to the pion mass, and in turn the thorium transition energy, at quadratic order in the axion field, in Appendix~\ref{ap:axion}.} to the SM by way of non-derivative operators.  We assume that this field comprises an  $\mathcal{O}(1)$ fraction\footnote{An exploration of the ability of scalar couplings to generate the correct relic abundance via thermal misalignment can be found in Ref.~\cite{Cyncynates:2024bxw}.} of the local DM abundance.  Working in units in which $\hbar = c = 1$ and following the notational conventions\footnote{In this paramaterisation, the coefficients and normalisation of each of the terms in the Lagrangian are specifically chosen such that the dimensionless couplings are all RG-invariant \cite{Damour:2010rp}.    } of Refs.~\cite{Damour:2010rp,Arvanitaki:2014faa} in parameterising the dimensionless couplings  relative to  Newton's gravitational constant $G_N$, the leading-order low energy Lagrangian, applicable to scales $\gtrsim$ 1 GeV, can be expressed in terms of a relatively minimal set of operators as
\begin{equation} 
\label{eq:lagrangian}
\begin{split}
\mathcal{L}_{\phi} = - \kappa \phi  \Biggr[ \frac{d_e}{4e^2} F_{\mu \nu} F^{\mu \nu} -  \frac{d_g \beta_3}{2 g_3} G^{A}_{\mu \nu} G^{A \mu \nu}  - d_{m_e} m_e \bar{e}e   \\ - \sum_{i = u, d} \left( d_{m_i} + \gamma_{m_i} d_g \right) m_i \bar{\psi}_i \psi_i \Biggr] ~,
\end{split}
\end{equation}
where $F_{\mu\nu}$, $G_{\mu\nu}$ are respectively the photon and gluon field strength tensors  and $e$, $\psi_i$ are the fermion fields of the first generation. 
Here, $\kappa = \sqrt{4 \pi G_N}$, $\beta_3 (g_3)$ is the QCD beta function, $g_3$ is the strong coupling constant and the $\gamma_{m_i}$, $i = \{u,d\}$ are the anomalous dimensions of the $u$ and $d$ quarks.

These effective operators feature in several well-motivated models of ultra-light dark matter. For one example, in the presence of additional sources of CP violation beyond those included within the SM, both the QCD axion, and other pseudo-scalar axion-like particles possess CP-even scalar couplings in addition to their CP-odd pseudo-scalar interactions. In particular, the QCD axion can attain a non-zero scalar coupling to quarks \cite{Pospelov:1997uv,Pospelov:2001ys,Pospelov:2005pr}, which is bounded from below by the CP violation in the SM, and from above by measurements of the neutron electric dipole moment. 

Owing to the high occupation numbers and small velocities  of DM in the Milky Way, $\phi$ can be  well approximated by a non-relativistic temporally and spatially oscillating classical plane wave
\begin{equation}
\phi(t,\mathbf{x}) = \phi_0 \cos[m_\phi (t - \mathbf{v\cdot x}) + \theta ] + \mathcal{O}(|\mathbf{v}|^2)~,
\end{equation}
with an amplitude $\phi_0 \simeq \sqrt{2\rho_{\textnormal{DM}}}/m_\phi$ which is set by the requirement that the field saturates the local DM energy density, $\rho_{\textnormal{DM}} = 0.3$ GeV cm$^{-3}$ \cite{Read:2014qva}. $\theta$ denotes an arbitrary phase. The field is temporally coherent over a coherence time $\tau_c = 2 \pi /( m_{\phi}\sigma_v^2 )$ and spatially coherent over a coherence length $\lambda_c = 2\pi / (m_{\phi}\sigma_v)$ \cite{Centers:2019dyn}  where $\sigma_v \sim \sqrt{\langle (\mathbf{v} - \langle \mathbf{v} \rangle )^2 \rangle} \sim 5 \times 10^{-4}$ (in natural units)  \cite{Evans:2018bqy} is the  DM virial velocity. 

\subsection{The signal}
\label{sec:sig}
The couplings in Eq.~\ref{eq:lagrangian} cause the fine structure constant $\alpha$, the strong coupling constant $\alpha_s$ and the fermion masses to vary in time.  Explicitly,
\begin{align}
m_\psi &= m_\psi \left[ 1 + d_{m_\psi}\kappa\phi(t,\mathbf{x}) \right] \\
\alpha &= \alpha \left[ 1 + d_{e}\kappa \phi(t,\mathbf{x}) \right] \\
\alpha_s &= \alpha_s \left[ 1 + d_{g}\kappa \phi(t,\mathbf{x}) \right]
\end{align}
where $\psi$ represents a generic fermion with mass $m_{\psi}$. 

These variations induce oscillations in the nuclear energy and length scales and, in turn, the thorium nuclear transition energy, $\omega_N = 8.3$\,eV \cite{Seiferle:2019fbe}.  $\omega_N$ is a function, $g$, of both $\alpha$ and the ratio $X_q = m_q / \Lambda_{\textnormal{QCD}}$, i.e.
\begin{equation}
\omega_N  = g(\alpha, X_q) ~~.
\end{equation}
Here, the QCD scale $\Lambda_{\textnormal{QCD}}$,  is defined to be the Landau pole of the running of the strong coupling constant as a function of an energy scale $\mu$, $\alpha_s(\mu) \sim 1/\ln(\Lambda_{\textnormal{QCD}}/\mu)$, and is thus related to $\alpha_s$ via dimensional transmutation. 

\begin{figure}
\centering
\includegraphics[width=0.4\textwidth]{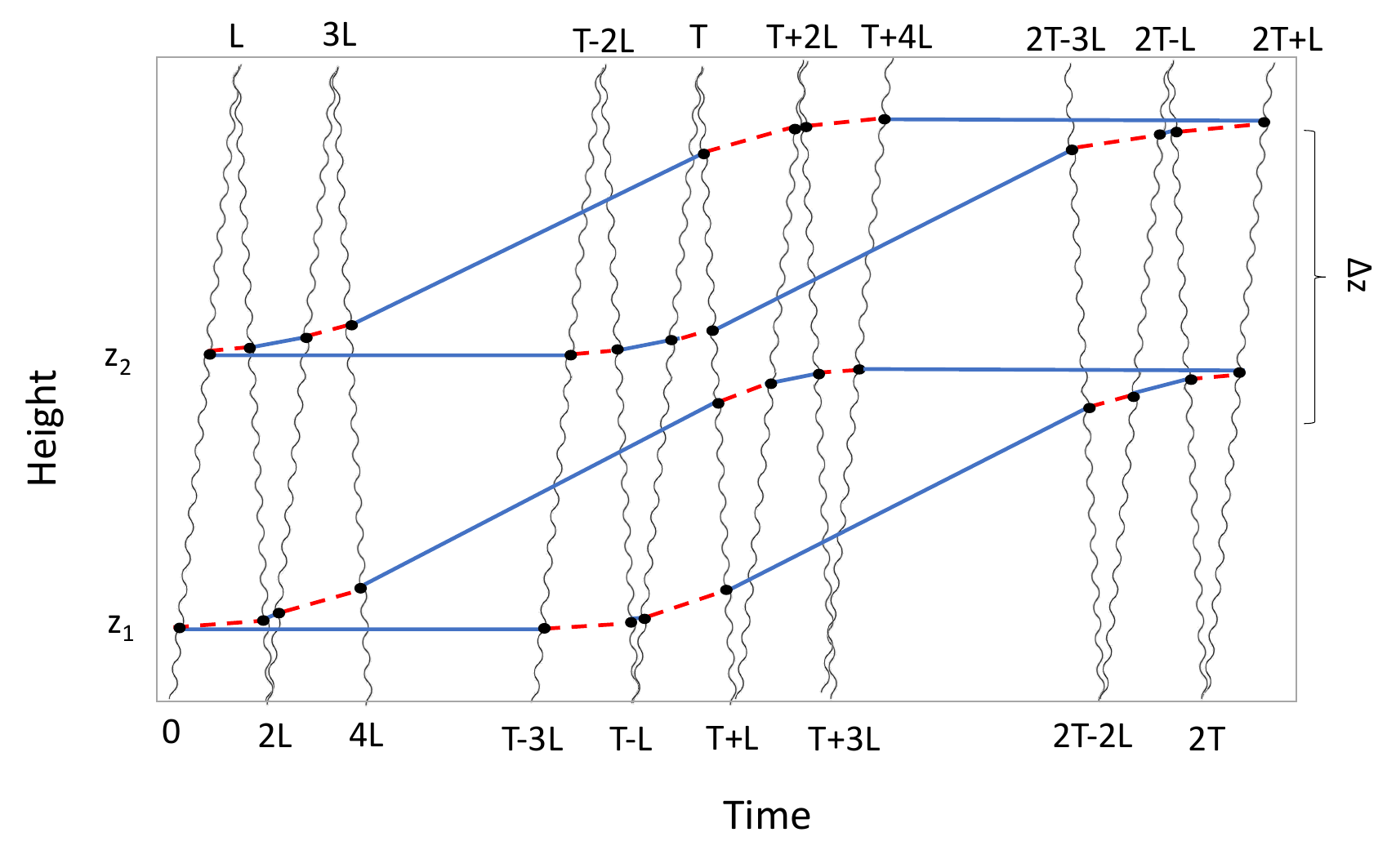}
\caption{A schematic spacetime diagram of a vertical single-photon atom gradiometer with $n = 4$ large-momentum-transfer atom optics. The experiment comprises a pair of single-photon atom interferometers based on single-photon transitions between a ground (blue) and excited (red) state. The wavy lines denote laser pulses fired from opposite ends of the baseline which are used to divide, direct and recombine the atomic states to yield an interference pattern. Atom-laser interactions are shown as black circles.}
\label{fig:spacetime}
\end{figure}

As such, in the presence of the scalar DM background $\omega_N$ oscillates as
\begin{equation}
\omega_N = \omega_N + \Delta \omega_N~, 
\end{equation} 
where \cite{Berengut_2011,Fadeev:2020bgt}
\begin{align}
\Delta \omega_N &= \omega_N \left(\frac{ \partial \ln g}{\partial \ln \alpha} \frac{\Delta \alpha}{\alpha} + \frac{ \partial \ln g}{\partial \ln X_q }\frac{\Delta X_q}{X_q} \right) \\
 &\approx \omega_N \left( 10^4  d_e + 10^5 (d_{\hat{m}} - d_g) \right) \phi(t, \mathbf{x})~.
\end{align}
 In this expression $d_{\hat{m}}$ is defined to be the coupling to the symmetric combination of quark masses:
\begin{equation} \label{eq:dm}
d_{\hat{m}} \equiv \frac{d_{m_{d}} m_d + d_{m_{u}} m_u}{m_d + m_u}~.
\end{equation}
 We now investigate how these oscillations would manifest in a future nuclear interferometer experiment. Fig.~\ref{fig:spacetime} shows a space-time diagram of the interferometer sequence proposed  for a vertical single-photon  atom gradiometer composed of a pair of interferometers situated at positions $z_1$ and $z_2$ respectively.   This scheme deploys large-momentum-transfer (LMT) techniques \cite{PhysRevLett.85.4498}  in order to increase the momentum transfer between the two arms of each interferometer and in turn the sensitivity. In practice this involves realising the beam-splitter with $n > 1$ laser pulses which each impart a momentum $\hbar \mathbf{k}$ where  $\mathbf{k}$ denotes the photon wavevector, to one of the interferometer arms. 

 The phase difference between the two arms of each isolated interferometer at the end of the interferometer sequence receives contributions from multiple sources. In the absence of new physics, the phase accumulated on each of the two arms is expected to cancel due to the symmetry of the interferometric sequence. It suffices therefore to only track the ULDM-induced phase accumulated by the excited state relative to the ground state over each arm of the interferometer. 
 
For a segment of the trajectory between times $t_0$ and $t_1$ during which the system is in the nuclear excited state, the ULDM-induced phase that is accrued (relative to the ground state) is
\begin{equation}
\Phi_{t_0}^{t_1} \equiv \overline{\Delta \omega_N} \int_{t_0}^{t_1}  \textnormal{d}t \  \cos(m_\phi t + \theta)~,
\end{equation} 
where we have defined \begin{equation}
 \overline{\Delta \omega_N}  = \omega_N \kappa \phi_0 (10^4 d_e + 10^5(d_{\hat{m}}-d_g))~,
\label{eq:ampw}\end{equation}
to be the magnitude of $\Delta \omega_N$.


  
The total phase difference recorded by a single  interferometer follows from subtracting the total phase (relative to the ground state) that is accumulated  on each of the two arms during propagation in the excited state. 

The gradiometer signal, $\Phi_s$, then corresponds to the difference between the phase difference recorded by the interferometer located at  position $z_1$ and  that recorded by the interferometer at position $z_2$.
For the interferometric sequence depicted in Fig.~\ref{fig:spacetime}, in the limit that the launch velocity in the two interferometers is the same, the gradiometer signal can be shown (see Ref.~\cite{Badurina:2021lwr} for a derivation) to be
\begin{equation}
\begin{split}
\Phi_s \approx \frac{\Delta z}{L}\left(\left[ \Phi^{T + L}_{T - (n-1)L}- \Phi_{0}^{nL}\right]\right.  -  \\ 
\left.\left[\Phi^{2T + L}_{2T - (n-1)L} -  \Phi_{T}^{T+nL}\right] \right)~,
\end{split}
\end{equation}
where the interrogation time $T$ is half the time between the two beam-splitter sequences, and $\Delta z = | z_1 - z_2 |$ is the vertical separation of the two interferometers.\footnote{For very long-baseline gradiometers such as those considered in Ref.~\cite{Arvanitaki:2016fyj}, the approximation $\Delta z / L \sim 1$ is often adopted. In more compact gradiometers including some of those considered in this work, this factor can be an important correction and needs to be explicitly considered.} $n$ refers to the number of large-momentum-transfer (LMT) kicks applied, i.e.\ the total number of pulses used to realise each beam-splitter sequence. 

The signal amplitude $\overline{\Phi}_s$ for a single phase measurement, defined according to 
\begin{equation}
\overline{\Phi}_s \equiv \left( \frac{1}{\pi} \int_0^{2\pi} \textnormal{d}\theta  \hspace{0.5em} \Phi_s^2 \right)^{1/2}~,
 \end{equation}
can then be shown to be~\cite{Badurina:2021lwr} 
\begin{equation}
\begin{split}
\overline{\Phi}_s = 8 \frac{\overline{\Delta \omega_N}}{m_\phi} \frac{\Delta z}{L} \sin \left[ \frac{m_\phi (T - (n-1) L)}{2}\right]  \\ \times  \sin \left[ \frac{m_\phi n L}{2}\right]  \sin \left[ \frac{m_\phi T }{2}\right]~.
\end{split}
\label{eq:amp}
\end{equation}

In order to isolate the dependence of this signal on the scalar couplings we introduce the notation
\begin{equation}\label{eq:fac}
\overline{\Phi}_s = d_{\phi}\overline{\Phi}_R~,
\end{equation} where \begin{equation}
d_{\phi} = d_e + 10 (d_{\hat{m}} - d_g)~.
\end{equation}

\subsection{Experimental Sensitivity}
\label{sec:sens}
We now quantify the reach of a potential nuclear interferometer as an ULDM detector. 
In practice, over an experimental campaign of duration $ T_{\textnormal{int}}$,  a total  of $N$ measurements of the gradiometer phase difference will be performed at a regular sampling interval $\Delta t $  =$T_{\textnormal{int}} / N $, corresponding to a constant shot repetition rate. Such measurements will contain, in addition to any possible ULDM-induced signal, contributions from background noise sources.  

Assuming that the sensitivity is limited by the quantum projection noise (or shot-noise) of the interrogated species with spectral density $S_{\textnormal{n}}$, the minimum scalar coupling $d_\phi^*$ that could be detected at a signal-to-noise ratio SNR following a measurement period $T_{\textnormal{int}}$ is
\begin{equation}
d_{\phi}^* \simeq \frac{\sqrt{\textnormal{SNR}}}{\overline{\Phi}_R} \times \sqrt{ \frac{S_{\textnormal{n}}}{T_{\textnormal{int}}}\times \textnormal{max}\left(1,\sqrt{\frac{T_{\textnormal{int}}}{\tau_c}}\right)}~.
\label{eq:sens}
\end{equation} Further details regarding the origin of this expression are given in Appendix~\ref{ap:uldmsens}.

Inserting Eq.~\ref{eq:amp} and Eq.~\ref{eq:fac} into Eq.~\ref{eq:sens}, one finds that the experimental sensitivity depends on the parameters $T,$ $L$, $n$ which are set in the experimental design, in addition to the shot-noise spectral density $S_{\textnormal{n}}$ and the campaign duration $T_{\textnormal{int}}$. For a gradiometer comprising two identical interferometers, the shot-noise can be described by the (frequency independent) white noise spectrum \cite{Le_Gou_t_2008,PhysRevA.47.3554,Badurina:2021lwr}
\begin{equation}
S_{\textnormal{n}} = \frac{2\Delta t}{C^2 N_\textnormal{s}}~,
\label{eq:noise}
\end{equation}
where $N_\textnormal{s}$ is the number of atoms (ions) per shot and the interferometer contrast, $C \leq 1$, characterises the amplitude of the oscillations of the probability of being in the ground and excited states. 

The values that these parameters can take depend both on the physical nature of the source (i.e. whether one is working with atoms or ions) and on the properties of the transition, including, most importantly for the scenarios to be considered in this work, its width. In clock interferometry, for which spontaneous decay of the excited state can be neglected,  $S_\textnormal{n}$ is independent of the choice of $T,$ $n$ and $L$ for a given set of starting conditions (i.e.\ the number of species per shot). Given the increase in the signal phase difference with $n,$ $L$ and $T$, such experiments look to maximise these parameters. Due to the geometric requirement that the interrogated clouds remain within the baseline for the duration of the interferometric sequence, it is important to realise that these parameters cannot be extremised independently. Any given choice of $L$ restricts the maximum LMT order $n$ and interrogation time $T$ that can be implemented without the trajectories of the interrogated species exceeding the baseline. 

The picture is more nuanced if one considers operating on a broad transition for which spontaneous decay becomes important relative to the proposed interferometer scales. Here, the increase in the phase shift that comes with increasing $L$ and $n$ is accompanied by a  degradation of the interferometer contrast and the number of species in the cloud at the point of detection relative to their input design values due to the enhanced degree of spontaneous decay. Precisely which of these two parameters is degraded depends on where in the interferometric cycle the decay occurs and  how far the decayed species have travelled away from the cloud before the end of the sequence. Those that have left the cloud contribute to a reduction in $N_\textnormal{s}$ whereas those which have lost coherence with the cloud due to decay but have not had time to travel significantly, form a diffuse background reducing $C$ \cite{sugarbaker2014atom}. Both of these effects manifest as an exponential increase in the shot-noise spectrum with the product $nL$ which, as we shall see, ultimately limits both the baseline and the LMT order that can be usefully deployed in such experiments.

 Before moving on we comment on the assumption that the experimental sensitivity is shot-noise limited. Achieving shot-noise limited detection is a key design goal for any atomic sensor and requires, by definition, that all other sources of noise arising from both systematic and stochastic effects, are subdominant. The use of a gradiometer configuration (as opposed to using a single interferometer) is key in this respect, allowing common mode noise which acts identically on the upper and lower interferometers, such as laser phase noise or mechanical jitters, to be rejected with a differential measurement. Whilst shot-noise limited sensitivity has already been demonstrated experimentally in table-top atom interferometers   (see e.g. Refs.~\cite{PhysRevLett.82.4619,89a3421464ae4f86825d63b2f9954141,Muller:2007es}),  scaling up to longer baselines brings additional considerations and challenges. In particular, coherence of the interrogated species needs to be maintained over the increased interferometry sequence duration which, for the instruments in question, is typically on the order of seconds. Comprehensive assessments of the many  sources of noise relevant to  long-baseline atom interferometry (e.g. gravity gradients, mechanical vibrations, black-body radiation etc.), in addition to various practices and techniques that may be deployed to control, or post-correct for these, have been discussed at length in literature (see e.g. Refs.~\cite{Dimopoulos:2008sv,Dimopoulos:2008hx,Abe_2021,Mitchell:2022zbp}) and used to place requirements on various aspects of the experimental operation. Where not already achievable with current technologies, these stand as targets for ongoing research and development. Given that the arguments for many of these noise sources carry over straightforwardly to both of the  nuclear interferometry realisations presented in this work, we do not attempt to replicate this discussion, assuming that such conditions and practices could also be successfully applied to any future nuclear interferometer experiment. We instead focus on those  aspects which are unique to the nuclear interferometer configurations presented here.  In  Sec.~\ref{sec:real}  we identify the most significant sources of noise likely to impact each of the proposals outlined in this work, compare them to the shot-noise floor  and discuss any resulting implications for the instrument design or experimental operation.

\section{Possible Realisations}
\label{sec:real}
We now discuss two distinct ways in which the principles outlined in the previous sections could be applied to perform interferometry on the ground-excited isomeric state transition in $^{229}$Th in the future  - the first involving single-ion `clocks', and the second clouds of neutral atoms. Both scenarios present a distinct set of experimental challenges from the sources traditionally considered for long-baseline interferometry, yet, as will be shown, may offer the potential to open up new phenomenological territory in the future provided the requisite technologies can be developed. Differing both in their physical nature and the properties of the ground-excited state transition, these two realisations demand starkly different operating configurations (as characterised by the parameters identified in Sec.~\ref{sec:sens}) which we now outline.  

\subsection{Single Ions}
As previously discussed, conventional ULDM searches using atom interferometry propose to  make use of ultra-narrow clock transitions which, thanks to long excited state lifetimes, allow for the use of long baselines and large inter-arm momentum transfers. Although the excited isomeric state lifetime for \textit{neutral} $^{229}$Th is only $\sim 10 ^{-5}$\,s \cite{Seiferle:2017lya} due to internal conversion, when ionised, this decay channel becomes inaccessible and the  (now radiative decay limited) lifetime is of order $\sim 10^4$\,s \cite{Kraemer:2022gpi,Tkalya:2015xia}, falling within the remits of clock interferometry.  We thus begin our exploration of the potential usage of $^{229}$Th in free-fall matter-wave interferometry by considering working with an ionic\footnote{From the perspective of the interferometry sequence, the precise charge of the ion is unimportant and we thus leave this unspecified.} form.                                                                                                            

Unlike for a neutral species which could (providing sufficient cooling is achieved to prevent significant Doppler expansion of the cloud) potentially be launched in clouds comprising up to $10^8 - 10^{10}$  simultaneously interrogated atoms, with ions electrostatic repulsion prohibits more than a single ion per shot (i.e.\ $N_\textnormal{s}$ = 1). Whilst, at least when exploited for fundamental physics, atom interferometers typically seek to maximise the number of atoms per shot in an effort to minimise quantum projection noise, experiments working with single atoms in free-fall have already been both conceived and performed \cite{Parazzoli_2012}.

In addition to deploying a high number of atoms per shot, the sensitivity forecasts for long-baseline atom interferometers also make use of multiplexing (see e.g. Ref.~\cite{Proceedings:2023mkp}). Exploiting the Doppler shift to the transition frequency experienced by clouds of atoms moving at different velocities, this technique allows the time interval between successive shot launches to be reduced below  the interferometer period (2T) with multiple independent interferometers concurrently in operation. Since (in typical Mach-Zehnder interferometry schemes) the atoms are launched upwards and subsequently fall under gravity, the trajectories of atoms in successive clouds cross. The atoms, which can be treated ballistically, are effectively non-interacting such that this trajectory overlap can occur without disrupting the generation of  interference at detection. The same can not be said for ions due to electrostatic repulsion. This imposes a lower limit on the time interval between shots:  $\Delta t \geq 2T$. We note that whilst this condition applies to interferometry configurations in which the ions are launched upwards and latterly fall under gravity as in this proposal, if instead a configuration could be utilised in which they only travelled in one direction (i.e. if dropped from rest with the primary laser pulses fired vertically downwards) such that they were detected in a different region from launch, it could be possible to achieve a higher effective shot flux due to the possibility of now applying multiplexing. In this case the experimental parameters would need to be carefully selected in order to ensure that there are no intersections between the trajectories of each of the arms of the two interferometers. In order for such linear, non-intersecting trajectories to simultaneously satisfy the geometric constraints of the baseline, smaller $T$, $n$ and $\Delta z$ than the  conventional set-up are likely to be necessary, somewhat suppressing the potential sensitivity gain associated with a higher ion flux. We leave a more detailed exploration of the potential benefit of such scenarios to future work.

In order to mitigate the inherently large quantum projection noise associated with working with single ions and thus optimise the sensitivity to ULDM of such an experiment, it is favourable to deploy both high LMT orders and long baselines. We emphasise that these practices are only possible thanks to the narrow transition linewidth in ionic form.   
In estimating the potential of  ion-based nuclear interferometry, we consider the possibility of both terrestrial and space-based experiments, taking guidance from the configurations forecast for the analogous future long-baseline clock atom interferometry experiments based on $^{87}$Sr (e.g. AION \cite{Badurina:2019hst}) in our selection of the relevant experimental parameters.  

For the terrestrial setup, we therefore consider a 1\,km baseline with $n$ = 1000 LMT optics. We set the interferometer separation $\Delta z$, to its maximum value 
\begin{equation}
\Delta z_{\textnormal{max}} \sim L - v_n T~,
\end{equation}
where \begin{equation}
v_n = n \omega_N / m_{\textnormal{Th}}~,\end{equation} is the velocity of the faster half of the wavepacket at the end of the beamsplitter sequence relative to the slower half and  $m_{\textnormal{Th}}$ denotes the mass of the thorium ion.
We select $T$ such that $\Delta z_{\textnormal{max}} \sim L$.

We model our space-based instrument on the design  considered in Refs. \cite{Graham:2016plp,PhysRevLett.110.171102,Graham:2017pmn,Tino:2019tkb,AEDGE:2019nxb}, comprising two satellites in medium Earth orbit separated by a distance of 4.4$\times 10^7$\,m along a single line-of-sight. We take the conservative assumption that the region over which the ions are interrogated is confined to within the interior of the satellites. This constrains the maximum wavepacket separation during the interferometric sequence, $v_n T$, to be less than $\sim$ 1\,m \cite{Graham:2017pmn,Graham:2016plp}.   Whilst this inequality is linear in both $T$ and $n$, we note that due to the dependence of the ion flux and thus the shot-noise spectral density on $T$, it is preferable to deploy larger $n$. We select $n$ and $T$ to satisfy this inequality, accounting for the fact that $n$ should be even and that the total duration of the  interferometer sequence should exceed the total time taken for the (sequentially fired) laser pulses to traverse the baseline, i.e. $2T > (4n-1)L/c$, where $c$ is the speed of light. Whilst the parameters used in our forecasts are representative of the reach that could be attained in such an experiment we note that further optimisation could be performed if one had a particular signal or region of parameter space in mind.  
Although more ambitious space-based designs in which the interrogation takes place exterior to the satellite allowing for wavepacket separations on the order of $\sim 100$\,m  have been proposed \cite{Dimopoulos:2008sv}, this removes the possibility of applying magnetic shielding and would thus leave the ions subject to uncontrollable electric and magnetic fields. This would likely be a major source of gradiometer phase noise which is exacerbated by the large interrogation times associated with this set-up\footnote{As will be discussed in due course, the gradiometer phase shift induced by electromagnetic field noise scales with $T^2$.} and we therefore deem such a design unsuitable and do not consider it further. For both the terrestrial and space-based designs we set $S_\textnormal{n}$ to be consistent with Eq.~\ref{eq:noise}, accounting for the fact that one must have both $N_{\textnormal{s}} = 1$ ions per shot and  $\Delta t \geq 2T $. 

 \begin{table*}
  \setlength{\tabcolsep}{10pt} 
\renewcommand{\arraystretch}{1.4} 
\begin{center}
\begin{tabular}{ |c||c|c|c|c|c| } 
 \hline
   Setup & $L$ [m]  & $T$ [s] &  $n$& $\sqrt{S}_{\textnormal{n}}$ [Hz$^{-1/2}$] & $\Delta z$ [m] \\ 
   \hline\hline
Terrestrial & 1000  &  1.2 & 1000& 2.3  & 986 \\
  \hline
  Space-based  & $4.4 \times 10^7$  &  5 & 16 & 4.7 & $4.4 \times 10^7$\\
  \hline

\end{tabular}
\caption{The experimental parameters characterising the single-ion nuclear interferometers considered in this work: the baseline $L$, the interrogation time $T$, the LMT order $n$, 
the shot-noise spectral density $S_{\textnormal{n}}$, and the vertical separation $\Delta z$ between the two interferometers. In the space-based configuration, the interrogation region is taken to be  within the interior of the  satellites.  }
\label{paramsion}
\end{center}
\end{table*}

With ions it is also necessary to factor in their enhanced sensitivity to  electromagnetic fields. Conventional atom interferometers already have stringent requirements on their internal magnetic environments in order to suppress unwanted contributions to the gradiometer phase which arise from (quadratic) Zeeman shifts to the transition frequency in the presence of temporally or spatially varying magnetic fields. For these interferometers there are two main sources of time-dependent magnetic fields that need to be considered: fluctuations in the local magnetic environment (i.e. for terrestrial interferometers the Earth's magnetic field), and variations in the bias magnetic field which is  applied to set a quantisation axis for the atoms and enable the use of magnetically insensitive states for which first order Zeeman shifts are not present. Whilst the use of a  bias field is not necessary for nuclear interferometry,\footnote{As will be discussed, the nuclear transition frequency is predicted to be largely insensitive to perturbations from external fields such that Zeeman shifts are strongly suppressed.} fluctuations in the environmental magnetic field remain a potential concern.

The transition frequency of proposed realisations of ionic nuclear clocks is predicted to be significantly more stable to perturbations from external fields than typical atomic optical clocks \cite{PhysRevLett.108.120802}. Instead, the dominant effect of an external magnetic field on the gradiometer phase arises due to perturbations to the ion's trajectory. For a single interferometer, the leading order  phase difference, $\Phi_i$, comes from the acceleration  of the interrogated species in the direction of the laser wavevector (i.e. the $z$ direction ):   $\Phi_i \sim n \omega_N a_z T^2/c$, where $a_z$ denotes the $z$-component of the acceleration experienced by the ions and $k$ is the magnitude of the laser wavevector \cite{hogan2008lightpulseatominterferometry,Dimopoulos:2008hx,PhysRevLett.110.171102}. A magnetic field of magnitude $B$ orientated perpendicularly to the $z$-axis  induces an additional vertical acceleration of   $a_z = qBv_t/m_{\textnormal{th}}$ on an ion of charge $q$ and mass $m_{\textnormal{th}}$, where $v_t$ denotes the speed of the particle in the transverse direction perpendicular to the magnetic field. Whilst, in the case of spatially and temporally uniform fields, the phase accrued by each isolated interferometer due to this effect is identical and thus cancels in a differential measurement, spatially dependent magnetic field fluctuations would produce spurious gradiometer phase noise, which, if located in the instrument frequency band, could be indistinguishable from an oscillating ULDM signal. 

In particular, if the two interferometers experience a difference in  magnetic field  of $\delta B$ in some direction perpendicular to the $z$ axis over the course of the interferometer sequence, an additional gradiometer phase shift of 
\begin{equation}
\begin{split}
\delta \Phi \sim \left(1\,\textnormal{Hz}^{-\frac{1}{2}}\right)  \left(\frac{q}{e}\right)\left(\frac{n}{1000}\right)\left(\frac{v_t}{0.1 \textnormal{mm/s}}\right) \times \\ \left(\frac{\delta B}{\textnormal{pT}/\sqrt{\textnormal{Hz}}}\right)\left(\frac{T}{ \textnormal{s}}\right)^2~,
\end{split}
\end{equation}
where $e$ = $1.6 \times 10^{-19}$\,C is the charge of the electron, will be induced.
Assuming the transverse velocities of the ions to be $\sim 0.1$\,mm/s corresponding to an effective temperature\footnote{Whilst techniques for cooling single trapped ions down to $\sim$ nK thermal temperatures have been proposed \cite{Cui:2023ajk}, we note that transverse velocities of $\sim 0.1$ mm/s could be realised at higher thermal temperatures by allowing for a greater longitudinal velocity. It is the use of such cooling practices in conventional atom interferometry which have given rise to the use of the notion of `effective' temperature - the thermal temperature  corresponding to the speed in the transverse plane.  }  on the order of  $\sim$ nK, the terrestrial and space-based configurations detailed in Tab.~\ref{paramsion} remain shot-noise limited for magnetic field fluctuations of amplitude $\delta B \leq 1$ $\textnormal{pT}/\sqrt{\textnormal{Hz}}$ and $\delta B \leq 7$ $\textnormal{pT}/\sqrt{\textnormal{Hz}}$, respectively. 
This expression assumes that the time scale over which the field fluctuates is long relative to the interrogation time such that the magnetic field experienced by each interferometer remains  approximately constant over the interferometer sequence, which is reasonable for the sub-Hz frequencies and $\mathcal{O}$(s) interrogation times of relevance here.  

It remains to compare these requirements to the magnetic field backgrounds that may be expected for these experiments. At present, the techniques for shielding magnetic fields over the large scales necessary for terrestrial very-long-baseline interferometry (i.e.\ 10's - 100's of metres) offer screening down to the $\sim$ nT level \cite{SCHMIDT1992569,Wodey:2019ges}.\footnote{Whilst the damping of magnetic field fluctuations and non-uniformities to sub-pT levels has been achieved in certain laboratory settings (see e.g.\ Ref.~\cite{Altarev:2015fra}), the shielding required for very-long-baseline interferometry is significantly more challenging owing both to the total shield length required \cite{Abe_2021} and to problems of flux leakage in large length-to-diameter ratio shields \cite{10.1063/1.4720943}.} The amplitudes of stochastic fluctuations in the terrestrial magnetic field at the surface of the Earth  in the frequency band of interest are expected to be below these levels \cite{2016SGeo...37...27C,CONSTABLE2023107090} and would therefore evade present technologies.  Whilst at frequencies exceeding  $\sim$ 1\,Hz, the magnitude of these fluctuations is sufficiently small to satisfy the condition for shot-noise domination, they rapidly increase in magnitude below this point \cite{2016SGeo...37...27C,CONSTABLE2023107090} such that in the absence of any other means of mitigation, the experiment would become magnetic noise limited with current technology. For the experiment to remain shot-noise limited across the entire operational frequency band, a three order of magnitude improvement in the shielding factor of large scale magnetic shielding would be required. Even without such advances however it may be possible to substantially reduce the impact of magnetic noise through active surveillance and mitigation of the local magnetic environment. To this end, it is possible to measure magnetic fields to sub pT levels with magnetometers. Using an array of such instruments to monitor the local magnetic field strength and direction  may allow for a reconstruction of the ion trajectories and the subsequent computation and subtraction of the corresponding gradiometer phase response, as outlined in Ref.~\cite{Dimopoulos:2008sv}. 

In contrast, temporal fluctuations in the interplanetary magnetic field (as relevant to a space-based instrument) of frequency $f$ at $\sim$1\,AU have been measured to be $\sim 0.1 \frac{\textnormal{nT}}{\sqrt{\textnormal{Hz}}} \times \left(\frac{10^{-2}\textnormal{Hz}}{f}\right) $ \cite{1969SoPh....8..155S}. If these fluctuations could not be shielded or damped, the proposed space-based nuclear interferometer would become limited by magnetic field noise at frequencies below $\sim 0.1$\,Hz. To this end, retaining shot-noise limited detection over the entire frequency band of the experiment demands either shielding capable of suppressing fluctuations down to the $\sim$ 7\,pT level across the $\sim$ 1\,m interrogation region, or a successful active monitoring and mitigation campaign \cite{Dimopoulos:2008sv}. Given the substantially smaller scales over which fluctuations need to be screened, meeting these shielding demands is likely to prove less challenging than those of the terrestrially-based case.  

Gradiometer phase noise can similarly arise from spurious vertical accelerations induced by  temporally fluctuating electric fields. Specifically, a difference in the $z$-component of the electric field experienced by each interferometer of $\delta E$ generates a gradiometer phase shift of 
\begin{equation}
\delta \Phi \sim  \left(1\,\textnormal{Hz}^{-\frac{1}{2}}\right) \left(\frac{q}{e}\right)  \left(\frac{n}{1000}\right) \left(\frac{\delta E}{\textnormal{0.1\,fVm}^{-1}/\sqrt{\textnormal{Hz}}}\right)\left(\frac{T}{\textnormal{s}}\right)^2~.
\end{equation}
Remaining in the shot-noise limited regime thus necessitates sufficient shielding of electrostatic fields to keep fluctuations below $\sim 0.1$ fVm$^{-1}/\sqrt{\textnormal{Hz}}$ and $\sim 0.7$ fVm$^{-1}/\sqrt{\textnormal{Hz}}$  in the terrestrial and space-based set-ups respectively. These requirements can be be translated to conditions on the metallic vacuum chamber in which these instruments would be housed.  

Local patch potentials, spatial and temporal variations of electrostatic potentials across surfaces \cite{vitale2024estimateforcenoiseelectrostatic}, which are responsible for short-range forces close to surfaces, may pose a further challenge. Given that the dominant way in which electric fields are imprinted in the gradiometer phase comes from  perturbations to the $z$-component of the acceleration of the ions, it is vertical E-fields, as generated from horizontal surfaces which are of greatest concern. In this respect, the ion trajectories (as determined by their launch heights and velocities) should be chosen to keep the ions sufficiently far from the surfaces at the upper and lower ends of the chamber.

\subsection{Neutral Atoms}
One could alternatively conceive a nuclear interferometer based upon the ground - isomeric excited state transition in \textit{neutral} $^{229}$Th atoms. Whilst this allows for significantly greater effective flux rates and thus much lower quantum projection noise, there are additional complexities relating to the short lifetime of the excited state which need to be carefully considered. Compared to clock interferometry, in the case of a broad transition the experimental sensitivity is negatively impacted by spontaneous decay occurring both during  periods of free propagation in the excited state and during atom-laser interactions. Each of these effects can be accounted for by the introduction of an appropriate correction factor as we shall now detail.

Within an order $n$ single-photon LMT interferometer sequence, the ensembles of atoms are required to spend a total of $n$ periods in the excited state, each lasting for a time $\sim 2L / c $  set by the time for light to travel across the baseline and back. The probability of an atom remaining coherently within the cloud at the end of the sequence is thus (reverting to physical units) \cite{Loriani:2018qej}
\begin{equation}
P_A = e^{\frac{-2nL}{c \tau}}~,
\end{equation}
where $\tau \sim 7\,\mu\textnormal{s} / \ln(2) \sim 10\,\mu$s is the (mean) lifetime of the neutral thorium isomeric excited state \cite{Seiferle:2017lya}. 

To avoid a significant deterioration of the sensitivity attributed to a given initial number of atoms,  the condition 
\begin{equation}
nL \leq \frac{c \tau}{2} \sim 1500~\textnormal{m}~,
\label{eq:limit} 
\end{equation}
should therefore be satisfied. Unlike single-ions, neutral $^{229}$Th is thus restricted to terrestrial scale interferometry experiments.

It is also important to account for the impact of spontaneous decay during the (finite duration) atom-laser interactions. This manifests as a reduction in the efficiency of the so-called $\pi$-pulses used to invert the populations of the ground and excited states. The proposed LMT interferometry scheme depicted in Fig.~\ref{fig:spacetime} relies upon a number of selective $\pi$-pulses which are designed to address only a single arm of the interferometer.\footnote{Other than the central pulse of the sequence which is a $\pi$-pulse that simultaneously acts on both arms, and the first and last pulses of the sequence which are $\pi/2$-pulses used to create a coherent superposition of ground and excited states, all of the pulses in Fig.~\ref{fig:spacetime} correspond to selective $\pi$-pulses whose frequencies are tuned to interact with a single arm.} As we shall now show, the requirement of selective addressing places an upper bound on the laser intensity that may be used to implement these pulses, and in turn a lower bound on their duration. For the $^{229}$Th transition considered here, the minimum  durations are comparable to the excited state lifetime, rendering spontaneous decay during the pulse a non-negligible effect. 

Consider the first $\pi$-pulse in the beam-splitter sequence i.e. the second pulse in Fig.~\ref{fig:spacetime}. At the time at which this pulse is applied, the atom is in a coherent superposition of ground and excited state wave packets, with the excited component (corresponding to the upper arm of the interferometer) moving upwards with speed $v_k=\hbar k/m_{\rm Th}$ relative to the lower arm. Here $m_{\rm Th}$ now denotes the mass of a $^{229}$Th \textit{atom}.  This first $\pi$-pulse propagates downwards  and is intended to induce stimulated emission in the upper arm (of each interferometer) whilst, crucially, leaving the lower arm unaffected. To conserve momentum, the upper arm recoils by a further $v_k$ in the $+z$ direction as a result of this process,  thereby increasing the velocity separation between the two arms.

Selective addressing is possible here because the desired stimulated-emission process, and the competing absorption process in the lower arm are resonant at different photon energies. The resonant energy of a transition is determined by the difference between the total energies of the initial and final atomic states, including both the internal-state energy and the kinetic energy associated with atomic recoil. For the pulse under consideration,  the desired and competing processes involve opposite changes in both the internal state and the atomic momentum and thus acquire different recoil-energy shifts. Consequently, the photon energy required to drive stimulated emission in the upper arm differs from that required to excite the lower arm. Explicitly this difference is
\begin{equation}
\Delta E_{1} \equiv \left | E_{1}^{\rm competing} -  E_{1}^{\rm desired} \right|  = \frac{2   k^2}{ m_{\rm Th}}~,
\label{eq:split}
\end{equation}
where the subscripts here denote that this is for the first $\pi$-pulse in the interferometry sequence, and we are once again working in natural units. This equation is derived explicitly in Appendix~\ref{app:deri}.  Similar expressions can be obtained for each subsequent selective $\pi$-pulse in the interferometry sequence by comparing the resonant energies of the corresponding desired and competing processes, carefully accounting for the direction of both the laser pulse and the atomic recoil. The resulting splittings are always an integer multiple $\ell$ of $\Delta E_1$. We henceforth classify $\pi$-pulses by their $\ell$-number and  denote the corresponding splitting between the desired and competing processes by $\Delta E^{\ell} \equiv \ell \Delta E_1$. For an $\ell$-type $\pi$ pulse to selectively address a single arm, it is necessary for $\Delta E^{\ell}$ to exceed the spectral width of the pulse, which we may approximate by its full-width-at-half maximum (FWHM)
 \begin{equation}
\delta E_{\rm laser} =  \Gamma_a \left(1 + \frac{I}{I_{\textnormal{sat}}}\right)^{1/2}~.
\end{equation} Here, 
$\Gamma_a = \tau ^{-1}$ corresponds to the natural linewidth of the ground-excited isomeric transition in atomic $^{229}$Th and the ratio of the applied laser intensity to the transition saturation intensity  $I/I_{\textnormal{sat}}$ captures the effect of power broadening due to the finite laser intensity. 

The condition 
\begin{equation}
\Delta E^{\ell} > \delta E_{\rm laser}~,
\label{eq:v}
\end{equation}
thus implies a maximum laser intensity for an $\ell$-type pulse of 
\begin{equation}
\frac{I}{I_\textnormal{sat}}\Bigg|_{\textnormal{max},\ell} = \left( \frac{2 \omega_N^2 \ell}{m_{\textnormal{Th}}\Gamma_a}\right)^2 - 1~,
\label{eq:Iratio}
\end{equation} where we have used that (in natural units) $k \sim \omega_N$.

Given that the Rabi frequency, $\Omega$, which sets the duration of the $\pi$-pulse according to 
\begin{equation}
\label{eq:pipul}
t_{\pi} = \frac{\pi}{\Omega}~,
\end{equation}
is related to $I/I_{\textnormal{sat}}$ by
\begin{equation}
\frac{\Omega^2}{\Gamma_a^2} = \frac{I}{2I_{\textnormal{sat}}}~,
\label{eq:omgam}
\end{equation}
Eq.~\ref{eq:Iratio} thus implies that there is a maximum achievable Rabi frequency which, by Eq.~\ref{eq:pipul}, translates to a lower bound on the duration of each selective $\pi$-pulse. Upon numerical evaluation we find that for $\ell = 1$, $t_{\pi,min} \sim 0.43 \tau$, rendering it important to account for spontaneous decay over the duration of the $\pi$-pulse. This can be formally achieved by explicitly including spontaneous decay in the optical Bloch equations which describe the behaviour of a 2-level system interacting with an electromagnetic field \cite{steck2007quantum, Scully_Zubairy_1997}. The result, which we detail  explicitly in Appendix~\ref{ap:pi}, is a reduction in the probability that an applied $\pi$-pulse results in population inversion from its ideal value of unity according to the value of the ratio $\Omega/\Gamma_a$. 

The interferometer sequence depicted in Fig.~\ref{fig:spacetime} involves (4$n$-3) $\pi$-pulses in total, all but one of which are intended to selectively interact with a single interferometer arm. For $n= 2 $, all four selective $\pi$-pulses are of $\ell=1$ type. For $n>3$ the sequence will also involve selective pulses of  higher $\ell$ in addition to eight $\ell=1$ pulses.  In estimating the total degradation of the sensitivity due to $\pi$-pulse inefficiencies over the course of the interferometer sequence, we assume that the laser intensity used for each $\pi$-pulse takes its maximum  value such that each pulse has as short a duration and thus high as efficiency as possible. In particular we note that for  $\ell = 1 $ type pulses the minimum pulse duration requires an intensity ratio of $I/I_{\textnormal{sat}}$ $\sim 105 $. The resulting pulse efficiency, obtained by evaluating Eq.~\ref{eq:piprob} at the maximum ratio of $\Omega/\Gamma_a$ according to Eqs.~\ref{eq:Iratio} and \ref{eq:omgam}, is  $P_1 = 0.85 $. 
Although increasing the LMT order $n$ yields larger momentum separations between the interferometer arms and hence a greater interferometric response, this comes at the cost of requiring additional selective $\pi$ pulses, each of which  induces substantial losses due to spontaneous decay. As a result, the total sequence efficiency falls rapidly with increasing $n$. For the $^{229}$Th transition this degradation more than offsets the benefit of larger momentum separations, making $n=2$ the optimal  choice.\footnote{Note that it is necessary to have $n > 1$ in order that the atoms predominantly propagate in the ground state allowing for interrogation times that exceed the excited state lifetime.} We therefore restrict our attention to $n=2$ sequences, which involve just four $\ell =1$ selective $\pi$-pulses, in addition to a central `mirror' $\pi$-pulse which is intended to simultaneously address both arms. Assuming that the intensity used for this central pulse is sufficiently high such that its duration is negligible with respect to $\tau$, the total probability of having undergone all the desired transitions at the end of an $n=2$ sequence is:
\begin{equation}
P_B = P_{1}^4~.
\end{equation}   

Combining the loss of coherence due to spontaneous decay  during free-propagation with the finite efficiency of the selective $\pi$-pulses, the net effect of spontaneous decay during the interferometer sequence is to increase the shot-noise spectrum to
\begin{equation} \label{eq:SN}
S_{\textnormal{n}}  = \frac{\tilde{S}_\textnormal{n}}{P_A P_B} =  \frac{ e^{\frac{2nL}{c \tau}} \tilde{S}_\textnormal{n}}{P_B}~,
\end{equation}
where here $\widetilde{S}_{\textnormal{n}}$ denotes the noise spectrum for an identical experiment subject to the same initial number of atoms per shot but in the absence of spontaneous decay. 

A crucial feature of  LMT interferometry is that the atoms propagate in the ground state for the majority of the sequence meaning that it is possible to make use of interrogation times which greatly exceed the excited state lifetime.  This has been experimentally demonstrated on table-top scales in Ref.~\cite{Rudolph:2019vcv} where LMT interferometry was performed on the 689\,nm intercombination line of $^{87}$Sr.\footnote{Note that the interferometer sequence deployed in Ref.~\cite{Rudolph:2019vcv} differs from that proposed in this work, in that every $\pi$-pulse is designed to address both arms of the interferometer simultaneously. This is realised by making use of sufficiently high laser intensities such that Eq.~\ref{eq:v} is not satisfied. } Whilst the short excited state lifetime does not therefore impose a direct upper limit on the duration of the interrogation sequence, the interrogation time does inherit a constraint indirectly from the geometric requirement that the atom trajectories remain within the baseline. This condition implies that for any given  choice of  $L$ and $n$ there is a maximum $T$ that can be utilised. In practice however, if looking to work in the regime $\Delta z \sim L$, it is necessary to deploy a value of $T$ substantially smaller than this maximum i.e.\ there is a trade-off between $T$ and $\Delta z$ which must be accounted for in  parameter selection.

In our sensitivity forecasts for this scenario, we consider three distinct operating configurations, characterised by the parameters $T,$ $L$ and $n$, the interferometer separation $\Delta z$ and the intrinsic phase noise $\sqrt{\tilde{S}_\textnormal{n}}$ that arises from the number of atoms launched per shot, representative of an `initial', `intermediate' and more `advanced'  experimental design. Whilst we specify our configurations using $\sqrt{\tilde{S}_\textnormal{n}}$, we emphasise that it is $S_\textnormal{n}$, as defined in Eq.~\ref{eq:SN}, which actually enters the sensitivity calculations in order to account for the additional loss of coherence/atom number due to spontaneous decay.  Some justification of these values, which we detail in Tab.~\ref{params_atom},  is in order.  As explained above, due to the substantial impact of spontaneous decay on the $\pi$-pulse efficiency we are restricted to an $n = 2$ LMT order. According to Eq.~\ref{eq:limit}, the greatest sensitivity will then be reached at a baseline of $\sim$ 750\,m which we thus deploy for our advanced configuration.  In order to take advantage of existing and planned experimental infrastructure, for the initial and intermediate configurations we specifically choose values for $L$ that match the baselines that are, or will be, deployed in the prototype iterations of proposed very-long-baseline clock interferometers based on the $^{87}$Sr clock transition.

For  $n = 2$, the maximum interferometer separation that can be achieved (assuming equal launch velocities) for a given $L$ and $T$ without the atom trajectories exceeding the baseline is approximately\footnote{We emphasise that this approximate expression is valid only for low ($\mathcal{O}(1)$) $n$. } 
 \begin{equation}
\Delta z_{\textnormal{max}} \sim L - \frac{1}{2}gT^2~.
 \end{equation}
Once again we set $\Delta z$ to $\Delta z_{\textnormal{max}}$ and choose $T$ such that $\Delta z_{\textnormal{max}} \sim L$. We explicitly include the $\Delta z / L $ factor in our sensitivity calculations, noting that this factor becomes increasingly important the shorter the baseline. 

As per Eq.~\ref{eq:noise}, in the shot-noise limit the phase sensitivity is essentially controlled at the design level by the time-averaged atomic flux ($N_{\textnormal{s}} / \Delta t $).  Given that atom interferometry relies upon the ensemble of atoms being sufficiently cold such that the fraction of atoms that are addressed with each laser pulse is not significantly degraded by the thermal expansion of the cloud over the course of the interferometer sequence, it is laser cooling techniques which ultimately limit the achievable flux rates. As discussed in detail in Refs.~\cite{Loriani:2018qej,dennis}, the percentage of atoms which interact with each laser pulse follows from a convolution of the spatial and velocity distributions of the ensemble of atoms with the laser pulse Fourier transform and spatial profile of the beam. In our forecasts we assume that the laser beam waist in use is sufficient to address every atom in every interaction. Since the clouds of atoms expand over time as a result of their thermal velocity, the initial temperature to which the atoms are cooled thus needs to be sufficiently low such that the cloud size remains appropriate for the beam waist in use. A greater understanding of the cooling temperatures and laser properties that may be required to achieve this could be developed through a more thorough analysis of the convolution integral. Assuming the usage of a similar beam waist, the requirements on the 3D momentum distribution of an ensemble of thorium atoms are likely to be of similar order of magnitude to those of more typical atom interferometry sources such as strontium (i.e. at the nK - pK level \cite{Dimopoulos:2008sv,Loriani:2018qej}). We thus inform our forecasts of potential fluxes with the initial and more ambitions projections discussed in the context of conventional atom interferometers (see e.g.\ Ref.\ \cite{Proceedings:2023mkp}). 

Whilst it is possible to produce clouds of $10^8 - 10^{10}$ atoms with current techniques \cite{metcalf1999laser}, the challenge in atom interferometry is to achieve the requisite level of cooling sufficiently rapidly so as to support high shot repetition rates and thus fluxes \cite{Dimopoulos:2008sv}.  The  spectral densities of $\sqrt{\tilde{S}_\textnormal{n}}$ = $10^{-4},$ $ 10^{-5} $ (Hz)$^{-1/2}$  projected for future long-baseline interferometers such as AION could (theoretically) be realised with $10^8$ and $10^{10}$  uncorrelated atoms per shot at a 1\,Hz shot repetition rate, however are in practice expected to be achieved with fewer atoms and the deployment of squeezing techniques in which the source atoms are entangled \cite{PhysRevLett.120.033601}.   In adopting these 
flux rates for our forecasts we are thus  implicitly assuming  the development of the requisite laser cooling and squeezing techniques for $^{229}$Th.  As is the case for conventional atom interferometers,  our projections should therefore  be interpreted as indications of the reach that could be achieved in the future should the necessary advancements in cold atom technologies be realised. On comparing our forecasts to those of conventional atom interferometry proposals however, it is important to keep in mind  that we have not accounted for any increased experimental challenge that may be associated with cooling and squeezing $^{229}$Th as opposed to $^{87}$Sr, and therefore that the timescales over which these technologies may be developed  are likely to differ. 

  \begin{table*}
  \setlength{\tabcolsep}{10pt} 
\renewcommand{\arraystretch}{1.4} 
\begin{center}
\begin{tabular}{ |c||c|c|c|c|c| } 
 \hline
   Setup & $L$ [m]  & $T$ [s] &  $n$& $\sqrt{\tilde{S}_{\textnormal{n}}}$ [Hz$^{-1/2}$] & $\Delta z$ [m] \\ 
   \hline\hline
 Initial & 10  &  0.6 & 2& 10$^{-4}$ & 8.2\\
  \hline
  Intermediate & 100  &  1.8 & 2& 10$^{-5}$ & 84\\
  \hline
Advanced &  750  & 3.1 & 2 & $0.3 \times 10^{-5}$ & 702\\
 \hline
\end{tabular}
\caption{The parameters characterising the atom-based nuclear interferometer configurations deployed in the sensitivity forecasts presented in this work. $\sqrt{\tilde{S}_{\textnormal{n}}}$ refers to the `intrinsic' phase noise set by the shot repetition rate and the number of atoms launched per shot. These values are corrected prior to entering our sensitivity estimates to account for the impact of spontaneous decay as detailed in the main text.  }
\label{params_atom}
\end{center}
\end{table*}

A second significant challenge of working with neutral $^{229}$Th is that  the 8.3\,eV  nuclear excitation energy is greater than the first ionisation potential of the atom\footnote{Note that this is not the case for ionic forms.} which is 6.3\,eV. With the interferometry schemes proposed in this work being reliant on the successful completion of a precise sequence of coherent, laser-induced nuclear excitations and de-excitations, it is necessary to explore the extent to which photo-ionisation would compete with these processes,  depleting the number of atoms at the point of detection and reducing the sensitivity. Whilst a first-principles evaluation of the photo-ionisation and photo-excitation rates of this system lies beyond the scope of this work, we offer a rough preliminary assessment here by comparing estimations of the photo-ionisation and nuclear photo-excitation cross-sections of atomic $^{229}$Th in literature. 

From the ground electronic configuration, a laser resonant with the isomeric  transition is sufficiently energetic to ionise the atom via removal of  electrons from either the 6d or 7s electronic shells. At leading order, these processes proceed via allowed (E1) electric dipole transitions. The  cross-sections for E1 photo-ionisation of the 6d and 7s electrons  are computed numerically in Ref.~\cite{PhysRevC.100.044306}  using the flexible atomic code (FAC) based upon the relativistic configuration interaction method~\cite{doi:10.1139/p07-197}. At an energy of $\sim$ 8\,eV these cross-sections are of order 10$^{-17}$ cm$^2$ \cite{PhysRevC.100.044306}.
In the low saturation limit of moderate laser intensities and a laser bandwidth $\Gamma_l$  that exceeds the natural linewidth of the transition, the resonance nuclear photo-excitation cross-section is given by \cite{vonderWense:2020kck}
\begin{equation}
\sigma_{\textnormal{ex.}} \sim  \frac{\lambda_0^2}{2\pi}\frac{\Gamma_\gamma}{\Gamma_l} \sim \left(3.6 \times 10^{-20}\right) \times \frac{\Gamma_\gamma}{\Gamma_l} ~\textnormal{cm}^2,
\end{equation}
where $\Gamma_{\gamma} \sim 10^{-9} \hspace{0.5em} \Gamma_{a}$ is the radiative decay rate and $\lambda_0$ is the transition wavelength. Setting $\Gamma_l$ to $\Gamma_a$, its minimum value consistent with the regime for which this equation is valid, gives $\sigma_{\textnormal{ex.}}\sim 10^{-20}$ cm$^2$, approximately three orders of magnitude below the photo-ionisation rate. Whilst, as stated in Appendix \ref{ap:pi}, our sensitivity forecasts assume the laser bandwidth to be narrow  relative to the natural linewidth to reduce atom loss from spontaneous decay, here we are just interested in an order of magnitude estimate of the nuclear excitation rate in order to assess the extent to which photo-ionisation is likely to be a problem.  Given that the nuclear excitation rate should be greatest in the low saturation limit, using this result provides a conservative assessment of the potential impact of photo-ionisation on the interferometry sequence and should thus suffice here.

These preliminary estimates indicate that, unless some means with which to suppress ionisation can be developed, performing interferometry with neutral thorium is unlikely to be feasible due to a significant depletion of the number of atoms remaining coherently in the cloud at detection. Given that the cross-sections of higher order (e.g. first forbidden magnetic dipole M1) ionisation processes are, according to the computations in Ref.~\cite{PhysRevC.100.044306}, approximately 5 orders of magnitude smaller than estimations of the nuclear photo-excitation rate, it should suffice to suppress just the two leading order E1 ionisation channels identified previously. To this end it may be possible to exploit arguments based on transition selection rules to achieve some degree of suppression by working with polarised clouds of atoms and making use of a careful choice of laser polarisation. A number of experimental studies have observed that the total photo-ionisation cross-section of polarised atoms depends on the relative orientation of the atom and laser polarisation (see Refs.~\cite{PhysRevLett.23.211, PhysRevLett.26.1416,Li_2006,Thini_2020}).  However, determining the optimal configuration, in addition to assessing the extent of suppression that may result, demands a full atomic calculation, and falls beyond the remits of this study. It may be that such a practice is insufficient to reach the desired 3 order of magnitude suppression on its own, and that it is necessary to look to other more active methods to achieve this. Various such techniques have been the subject of both theoretical and experimental attention in literature (see Refs.~\cite{AMPopov_2003,Fedorov2006,PhysRevLett.65.2362,PhysRevA.69.033404,PhysRev.128.681,RevModPhys.40.441}) and have been shown capable of suppressing photo-ionisation by several orders of magnitude \cite{PhysRevLett.101.163002}. The assessment of whether or not such methods (or indeed others) could be adapted  to suppress photo-ionisation in  $^{229}$Th in a way that is compatible with the proposed interferometry sequence lies beyond the scope of this work. 

In light of this, in the forecasts presented in Sec.~\ref{sec:diss} of this work we show the reach that could be achieved should it be possible to develop appropriate techniques to suppress photo-ionisation to negligible levels. In this way our estimates serve as motivation from a fundamental physics perspective for a dedicated programme aimed at establishing the feasibility of, and subsequently developing, appropriate atom stabilising technology for $^{229}$Th.

Whereas the leading source of noise other than shot-noise affecting the aforementioned single-ion  proposal was magnetic field noise, given both the electrical neutrality of atoms and the insensitivity of the nuclear clock transition to perturbations from external fields, such fluctuations are much less of a concern here.  The attainment of shot-noise limited sensitivity in this case is instead challenged by so-called gravitational gradient noise (GGN)\footnote{Whilst GGN similarly affects the ionic proposal previously discussed, it is substantially smaller than the high shot-noise resulting from single ion shots and thus does not limit the sensitivity. } which arises due to mass density fluctuations of the ground and atmosphere around the experiment \cite{Harms:2015zma,PhysRevD.58.122002}. These fluctuations perturb the local gravitational potential experienced by the clouds of atoms, generating  spurious gradiometer phase shifts.  This type of noise cannot be shielded and has been identified as a major challenge for  conventional terrestrial very-long-baseline atom interferometry proposals \cite{Proceedings:2023mkp}. For typical 100\,m- and km-scale instruments, GGN is to expected surpass the shot-noise at $\sim$ sub-Hz frequencies, leading to a significant loss of sensitivity  \cite{Vetrano:2013qqa,PhysRevLett.108.211101,PhysRevD.88.122003,Mitchell:2022zbp,Badurina:2022ngn}. 

 The modelling and mitigation of GGN at terrestrial atom interferometers constitutes an active and involved area of research (see e.g.\ Ref.~\cite{Badurina:2022ngn}) which depends both on details of the interferometry sequence and the local seismic environment of the experiment. Given this latter consideration, we do not attempt to perform any first-principles calculations here, but instead estimate the potential impact of GGN on the atom-based nuclear interferometer designs specified in Tab.~\ref{params_atom} by adapting existing analyses for standard long-baseline atom interferometers presented in the literature  to account for differences in the interferometry sequence and instrument design. 

The GGN spectral density for both a 100\,m- and a km-scale atom interferometer with the design specifications detailed in Tab.~\ref{aion:params} were calculated in Ref.~\cite{Badurina:2022ngn} in the $10^{-3} - 10$\,Hz frequency band  by modelling the seismic field as an incoherent superposition of monochromatic plane waves  propagating isotropically at the Earth's surface. Results for both a representative low and high noise seismic environment  were presented, based respectively on the empirically driven Peterson new low and high noise models \cite{peterson2003observations}.

At low frequencies, the  GGN spectral noise density, $S_{\textnormal{ggn}}$,   scales approximately with the interferometer design parameters as $\sim n^2 L^2 T^4$ \cite{Badurina:2022ngn}. To obtain a conservative estimate the impact of GGN on the nuclear interferometer configurations  considered in this work, we re-scale the high noise $S_{\textnormal{ggn}}$ presented in Ref.~\cite{Badurina:2022ngn} according to the parameters of our instruments and locate the frequency at which this becomes dominant over the shot-noise.\footnote{We emphasise that it is $S_n$, the shot-noise spectral density accounting for spontaneous decay which should be used for this comparison.} Due to both the substantially lower  $n$ and the increased shot-noise (due to the corrections for spontaneous decay) of the intermediate and advanced nuclear interferometer designs proposed here,  we find that they remain shot-noise limited down to lower frequencies:  $\sim10^{-3}$\,Hz and $\sim 10^{-2}$\,Hz respectively, compared to the $\sim$ Hz threshold of the 100\,m and km instruments of Tab~\ref{aion:params}.

The analysis of Ref.~\cite{Badurina:2022ngn} found that the sensitivity of the 10\,m instrument defined in Tab.~\ref{aion:params}  was not impacted by GGN in the frequency band considered in either the low or high noise scenarios. It is therefore safe to assume that our 10\,m initial nuclear interferometer would also remain in the shot-noise limited regime above $10^{-3}$\,Hz, however a full calculation of the GGN spectral density at lower frequencies would be required to obtain a more precise handle on the frequency down to which this is the case. For this reason, when we delineate the frequencies at which different instruments may become GGN-limited on our plots in Sec.~\ref{sec:diss}, we conservatively take a $10^{-3}$\,Hz threshold for the 10\,m instruments of both  Tab.~\ref{params_atom} and Tab.~\ref{aion:params}.

Given the potential impact of GGN at terrestrial atom interferometers, various passive mitigation techniques, which may also be applicable to nuclear interferometry, are under active investigation. In particular it was shown in Ref.~~\cite{Badurina:2022ngn} that  multigradiometer configurations in which multiple interferometers are positioned along the same baseline, could greatly reduce the impact of GGN on ULDM searches and recover significant regions of parameter space accessible to shot-noise limited instruments. A carefully selected site location with a quiet seismic environment (i.e.\ approaching the Peterson low noise as opposed to high noise model) may also offer the opportunity to regain up to three orders of magnitude in sensitivity \cite{Proceedings:2023mkp}.

\section{Discussion}\label{sec:diss}
\begin{figure}
\centering
\includegraphics[width=0.45\textwidth]{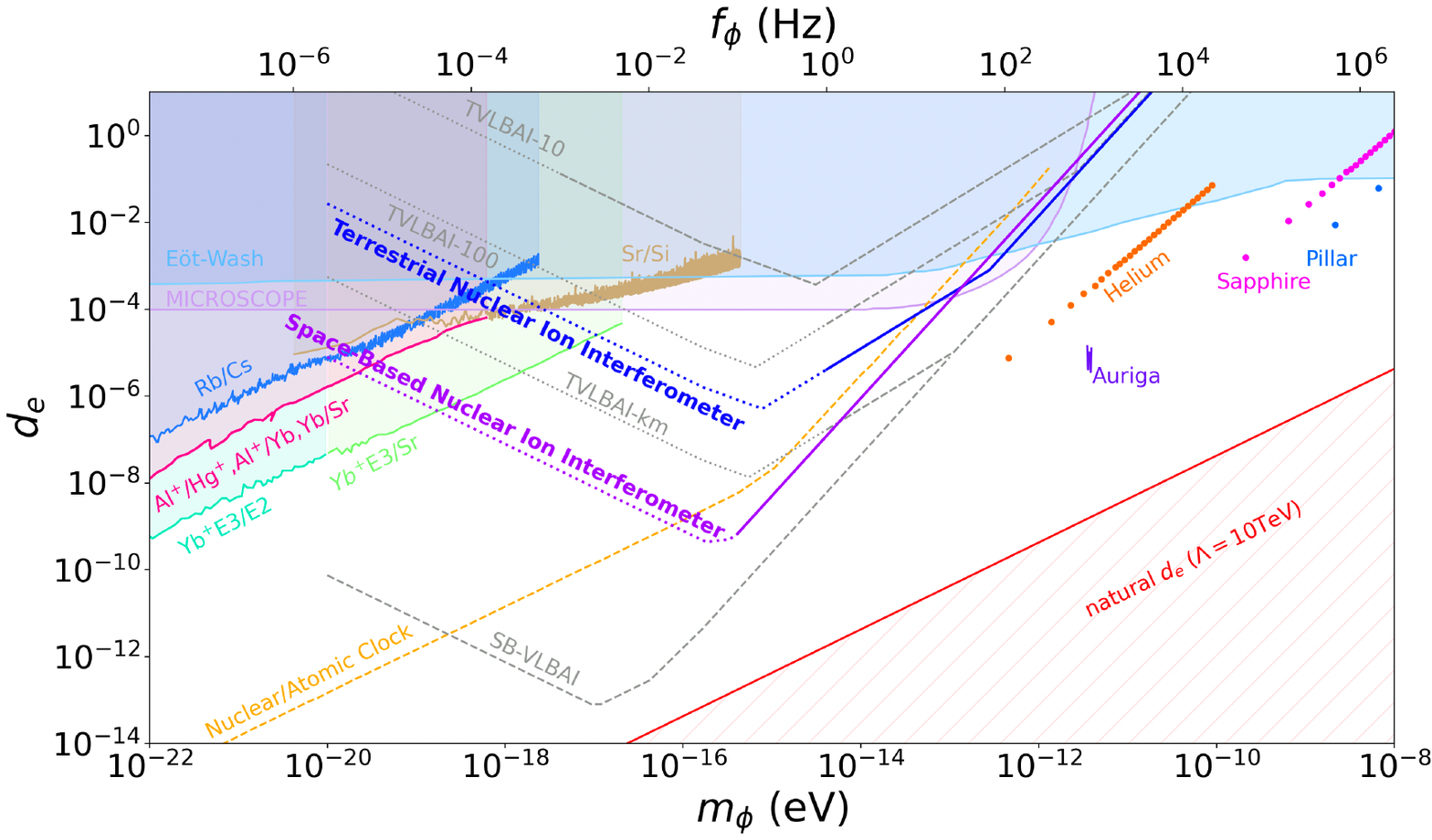}
\caption{Shot-noise limited projected sensitivity of the single-ion nuclear interferometer configurations detailed in Tab.~\ref{paramsion} to ULDM with a linear scalar coupling to photons. The curves show the minimum coupling $d_e$ that could be detected at a SNR = 1 following an experimental campaign of duration $T_{\textnormal{int}}=10^8$s. The dotted portion of the curve indicates where the sensitivity may be adversely impacted by magnetic field noise according to the discussion in Sec.~\ref{sec:real}. Also shown are existing bounds and forecast sensitivities of other experiments as detailed in the main text. The parameter space motivated from a naturalness perspective as described in the text is shown in red. }
\label{fig:deb}
\end{figure}

\begin{figure}
\centering
\includegraphics[width=0.45\textwidth]{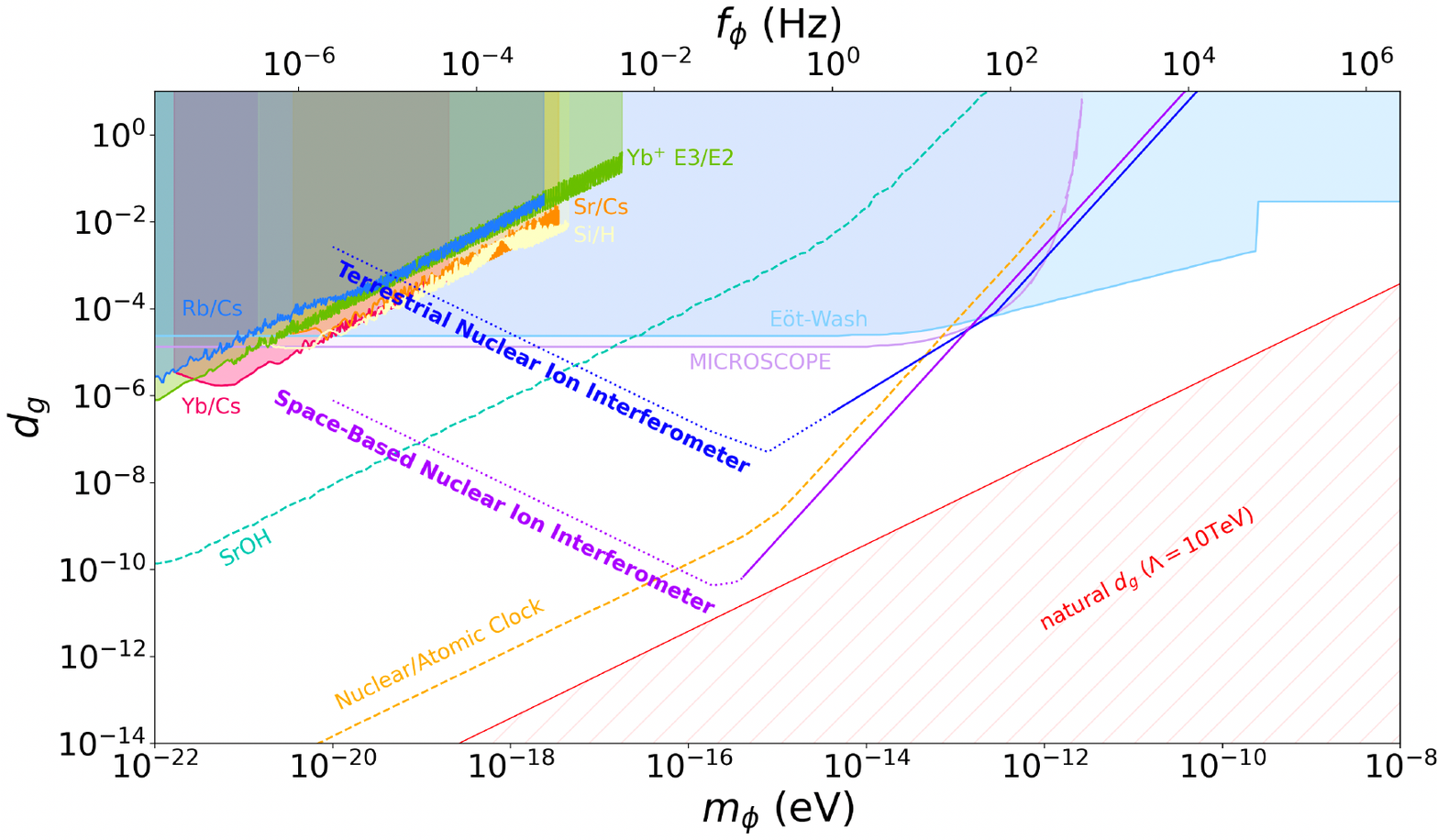}
\caption{Same as Fig.~\ref{fig:deb} but for ULDM with a linear scalar coupling to gluons. }
\label{fig:dgb}
\end{figure}

\begin{figure}
\centering
\includegraphics[width=0.45\textwidth]{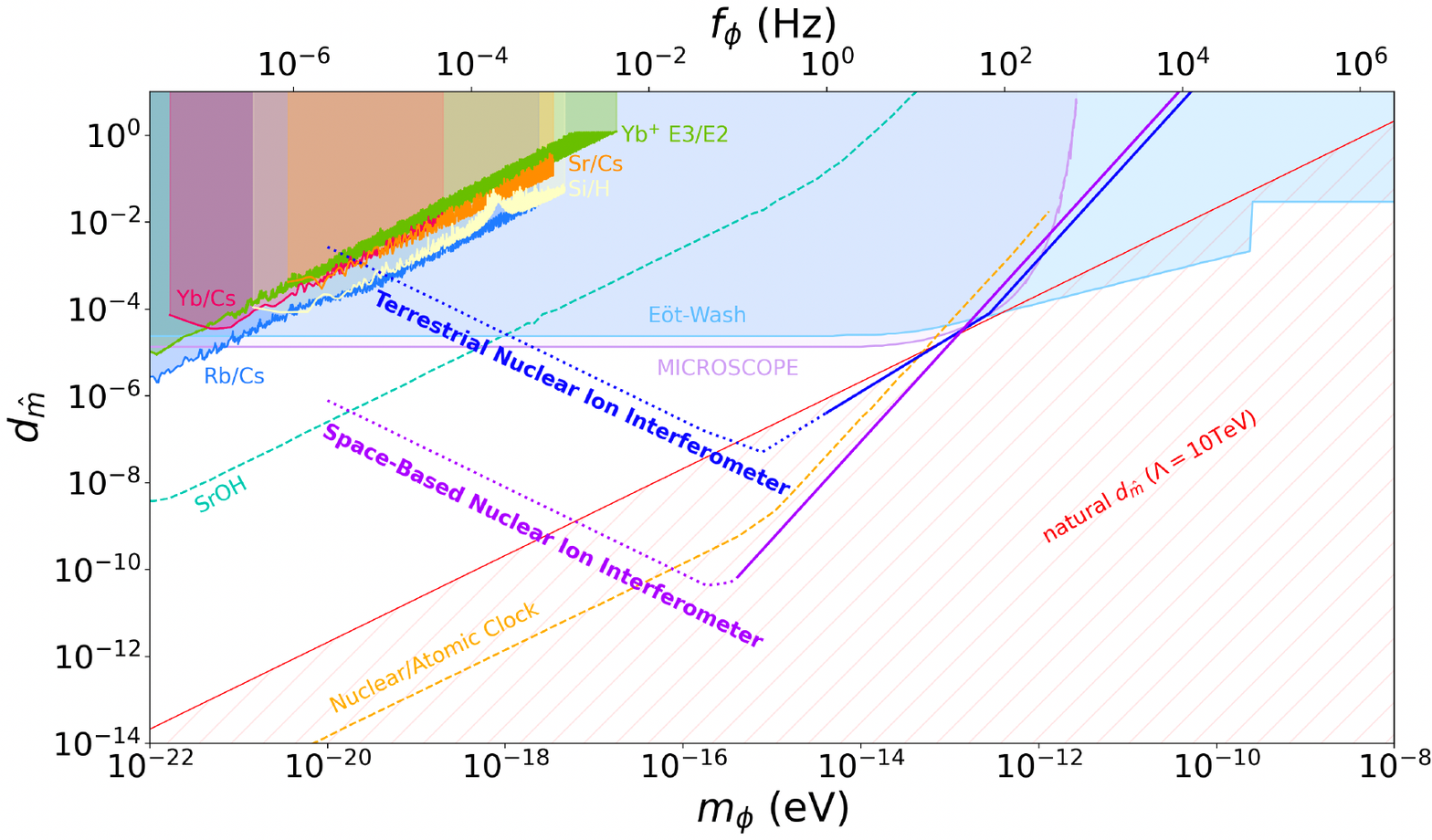}
\caption{Same as Fig.~\ref{fig:deb} but for ULDM with a linear scalar coupling to quarks.}
\label{fig:dmb}
\end{figure}

\begin{figure}
\centering
\includegraphics[width=0.45\textwidth]{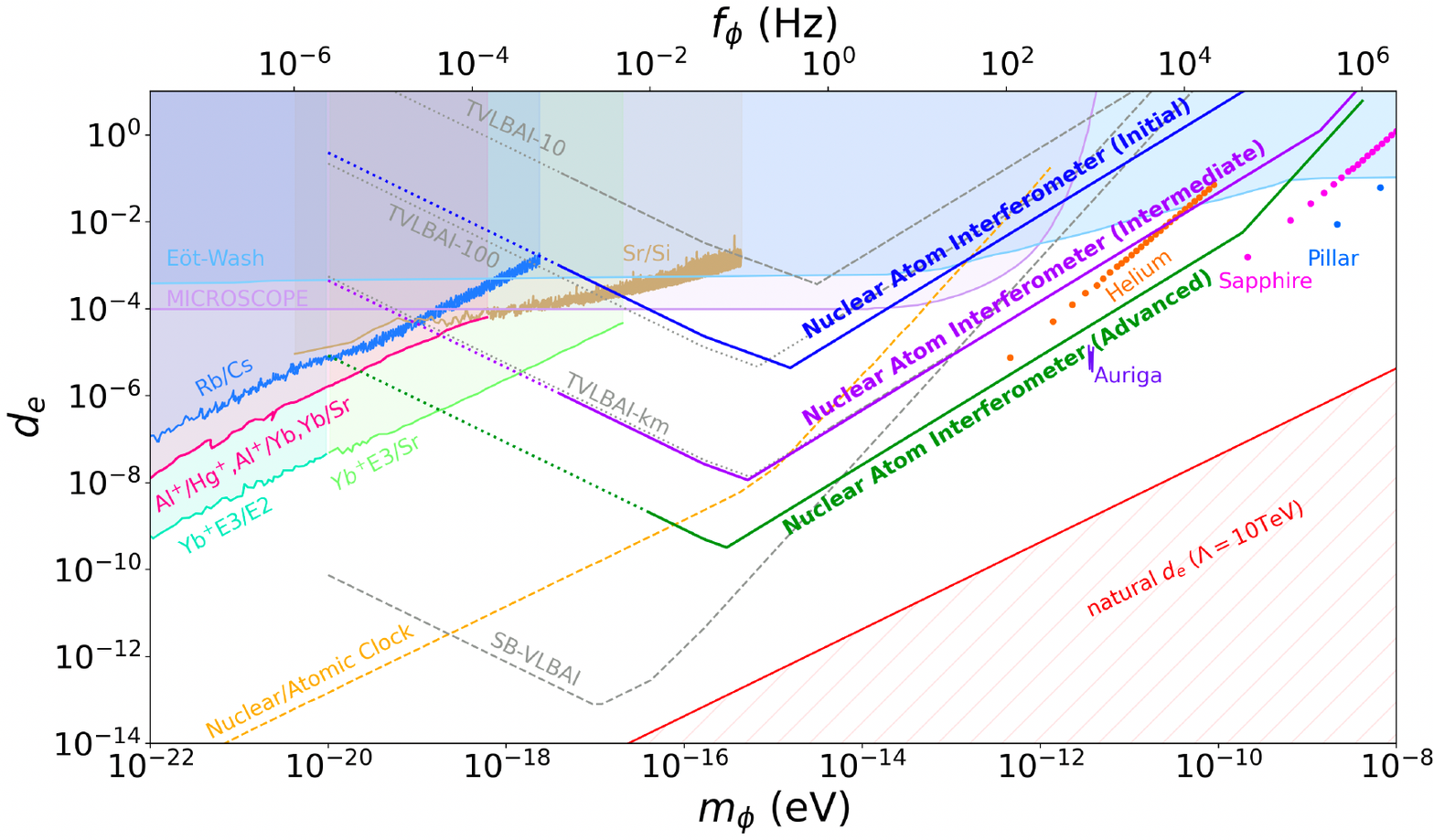}
\caption{Shot-noise limited projected sensitivity of the  nuclear interferometer configurations deploying neutral atoms detailed in Tab.~\ref{params_atom} to ULDM with a linear scalar coupling to photons. The curves show the minimum coupling $d_e$  that could be detected at a SNR = 1 following an experimental campaign of duration $T_{\textnormal{int}}=10^8$\,s and assuming that photo-ionisation has been successfully suppressed. The dotted portion of the curve indicates where the sensitivity may be adversely impacted by GGN according to the discussion in Sec.~\ref{sec:real}. Also shown are existing bounds and forecast sensitivities of other experiments as detailed in the main text. The parameter space motivated from a naturalness perspective as described in the text is shown in red. }
\label{fig:de}
\end{figure}

\begin{figure}
\centering
\includegraphics[width=0.45\textwidth]{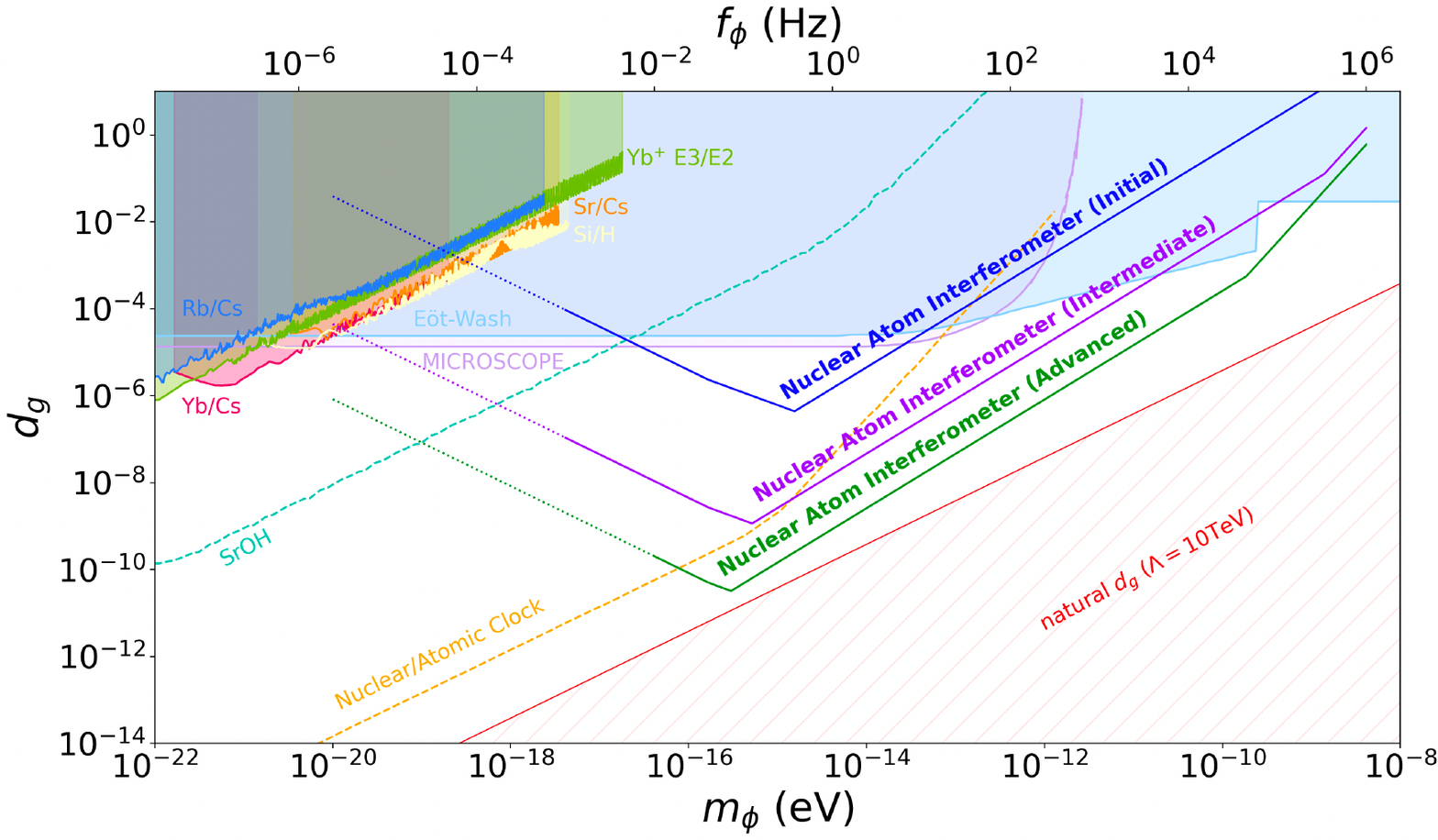}
\caption{Same as Fig.~\ref{fig:de}  but for ULDM with a linear scalar coupling to gluons. }
\label{fig:dg}
\end{figure}

\begin{figure}
\centering
\includegraphics[width=0.45\textwidth]{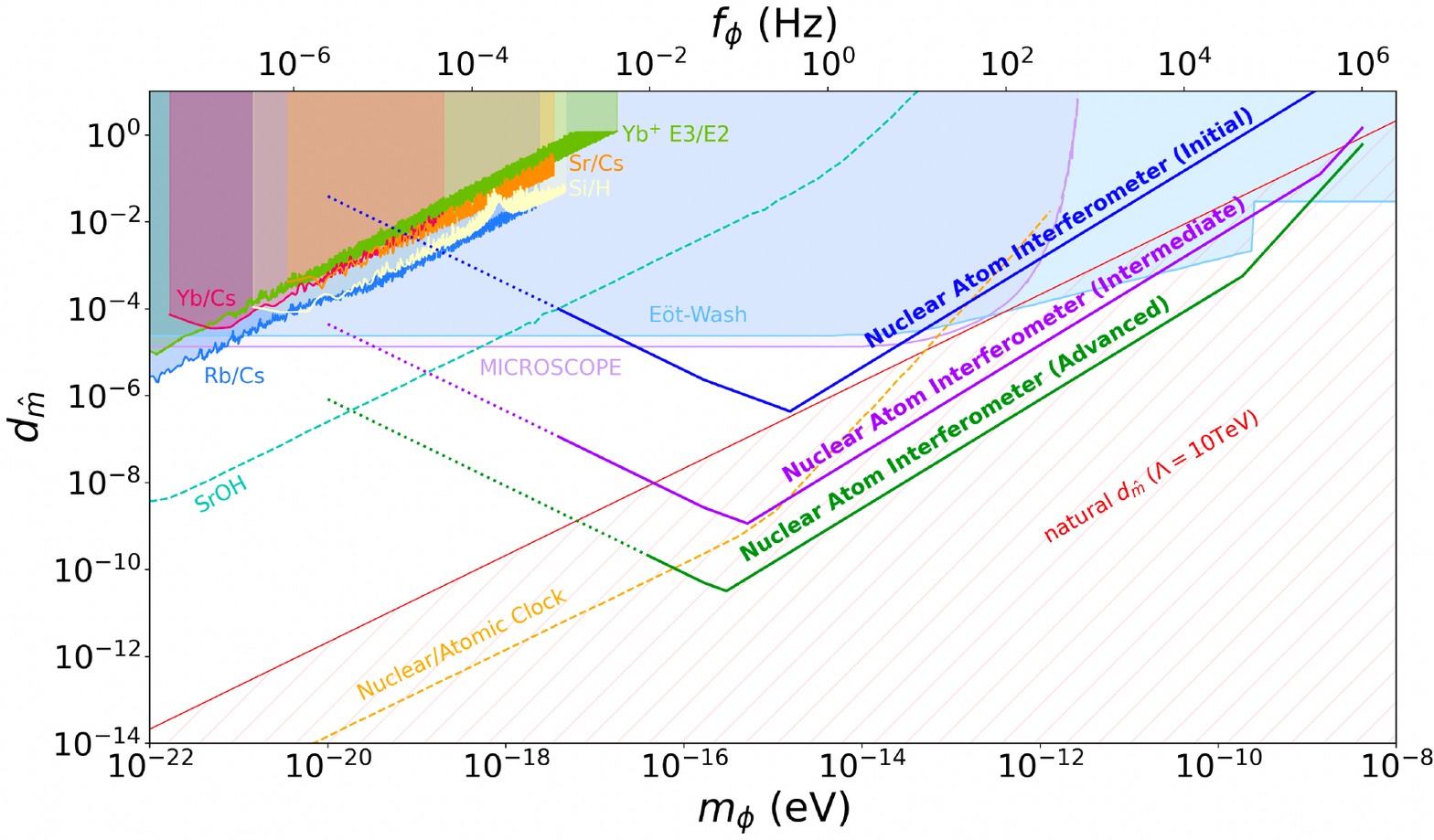}
\caption{Same as Fig.~\ref{fig:de} but for ULDM with a linear scalar coupling to quarks.}
\label{fig:dm}
\end{figure}

In Figs.~\ref{fig:deb}, \ref{fig:dgb} and \ref{fig:dmb}, we plot the couplings $d_e$, $d_g$ and $d_{\hat{m}}$ that could be detected at a signal-to-noise ratio (SNR) of 1 with the single-ion interferometer designs specified in Tab.~\ref{paramsion} following a measurement period of  $T_{\textnormal{int}} = 10^8$\,s, assuming that the coupling in question dominates $d_{\phi}$.  In Figs.\ \ref{fig:de}, \ref{fig:dg} and \ref{fig:dm}, we  present the shot-noise limited reach of the neutral atomic interferometer configurations of Tab.~\ref{params_atom},  once again taking   $T_{\textnormal{int}} = 10^8$\,s and assuming the suppression of photo-ionisation to negligible levels.  For clarity, our plots show the power-averaged envelope of the experimental response using the approximation  $\sin x$ = min($x$, $1/\sqrt{2}$) as in Ref.~\cite{Arvanitaki:2016fyj}. 

We emphasise that our forecasts assume the nuclear interferometer sensitivity to be shot-noise limited over the entire frequency band of the experiment. As discussed in Sec.~\ref{sec:real}, there may however be other sizeable noise sources which could become dominant. In order to reflect this, we indicate on our plots the frequencies, as determined by the analysis in Sec.~\ref{sec:real}, at which the experiment may depart from the shot-noise limited regime due to either magnetic field noise (in the case of ions) or GGN (in the case of atoms) with the use of a dotted, as opposed to a solid, line. We highlight that these thresholds assume that no attempts to mitigate either noise source either actively or passively are made. Advancements in large-scale magnetic shielding technology, a well-chosen site location and close environmental monitoring could all help to extend the shot-noise limited regime into lower frequencies and to recover regions of the impacted parameter space. 

For context, also shown in these figures are the 95\% confidence level (CL)  limits on the considered couplings from tests of the Equivalence Principle (EP)  both with  E\"{o}t-Wash torsion balance experiments \cite{PhysRevD.61.022001,Schlamminger:2007ht} and MICROSCOPE \cite{Touboul:2017grn,Berge:2017ovy}, as computed in Ref.~\cite{Hees:2018fpg}. We also display the limits from a hyperfine frequency comparison of rubidium and caesium (Rb/Cs) atomic microwave clocks \cite{Hees:2016gop}, in addition to the estimated reach of a frequency comparison between the $^{229}$Th nuclear clock and an atomic optical clock \cite{Arvanitaki:2014faa, Antypas:2022asj}. Upon the recommendation of Ref.~\cite{Centers:2019dyn}, the Rb/Cs limits derived in Ref.~\cite{Hees:2016gop} have been rescaled by a  factor of 3 in order to properly account for the stochastic nature of the scalar ULDM field and enable consistent comparison with competing bounds and proposals. 

In the case of the coupling to photons $d_e$, we also show limits from the AURIGA detector \cite{Branca:2016rez}, the frequency comparison of the electric octupole and the electric quadrupole clock transitions in single-ion ytterbium (Yb$^+$ E3/E2), the  comparison of the Yb$^+$ E3 transition to the strontium $S\to P$  clock transition (Yb$^+$ E3/Sr) ~\cite{Filzinger:2023zrs}, the comparison of a strontium optical clock with a silicon cavity (Sr/Si) \cite{Kennedy:2020bac} and the combined constraints of several optical clock comparisons (Al$^{+}$/Hg$^{+}$, Al$^{+}$/Yb, Yb/Sr) from the Boulder atomic clock optical network  \cite{BACON:2020ubh}. Forecasts for the sensitivities of proposed compact mechanical resonators based on superfluid helium, sapphire cylinders and sapphire pillars \cite{Manley:2019vxy} are also shown, as are the prospective shot-noise limited sensitivities of very-long-baseline atom interferometers operating on the $^{87}$Sr atomic clock transition. We show forecasts for terrestrial configurations (TVLBAIs) with 10\,m, 100\,m and 1\,km scale baselines in addition to a potential space-based realisation (SB-VLBAI), assuming the experimental parameters listed in Tab.~\ref{aion:params}. Further details on the calculation of these estimates, including a justification of the parameters used, can be found in Appendix~\ref{ap:aion}. As for the atom-based nuclear interferometers we indicate the frequencies at which the sensitivities of these experiments are likely to be impacted by GGN, assuming both a high seismic environment and that no mitigation strategies are applied, according to the analysis in Ref.~\cite{Badurina:2022ngn}, by switching from a dashed to a dotted line. 

In the case of the couplings to quarks and gluons, we show bounds from the comparison of a Yb lattice clock with a caesium fountain microwave clock (Yb/Cs) \cite{Kobayashi_2022}, the comparison of strontium  and  caesium clocks (Sr/Cs) \cite{Sherrill:2023zah}, the comparison of a hydrogen maser with a silicon cavity \cite{Kennedy:2020bac} and the comparison of the E3/E2 transitions in Yb$^{+}$ \cite{Banerjee:2023bjc}. Whilst the first three of these comparisons involve a microwave hyperfine transition, in the latter,  the sensitivity to $d_g$ and $d_{\hat{m}}$ derives from ULDM-induced oscillations of the nuclear charge radius. Also shown are projections for the comparison of a molecular SrOH clock with an atomic optical clock, assuming a total data accumulation period of 1\,year \cite{Kozyryev_2021}. 

We emphasise that whilst a variety of other experiments have probed, or have been proposed to probe each of these couplings, for clarity and efficiency, we have shown here just a representative selection of those which yield the strongest bounds or greatest prospective sensitivities in the parameter space of relevance. A more detailed overview of the landscape of competing experiments can be found (e.g.\ for $d_e$) in Ref.~\cite{Antypas:2022asj}.

The red shading in each of these figures delineates the parameter space which is motivated from a naturalness perspective.  We derive this following Ref.~\cite{Arvanitaki:2016fyj}.  The central argument here is that these scalar interactions break any shift symmetry which could keep the scalar light, such that a mass-squared contribution is expected from within the UV theory, scaling as the square of the coupling.  Motivated by the scales already explored by the LHC, we conservatively take the cutoff of the IR (infrared/ low-energy) theory at which these UV-contributions are predominantly generated to be $10$\,TeV and determine the parameter space in which these generic UV contributions are smaller than the scalar mass, thus avoiding the requirement of fine-tuning of different UV contributions to arrive at a given scalar mass, given a certain scalar coupling. Note, however, that this line is only intended as a rough guide.  Plausible UV completions could enter at lower scales, allowing for larger natural couplings.

It is important to realise that whilst the bounds from tests of the EP are independent of the abundance of $\phi$, the other experiments shown on these plots, including our forecasts for the various nuclear interferometer realisations, implicitly assume $\phi$ to comprise the totality of the relic local DM abundance. At low masses $m_{\phi} < 2 \times 10^{-20}$ eV, this scenario has been ruled out by examination of  small-scale cosmological structure in the Lyman-$\alpha$ forest \cite{Rogers:2020ltq}.

Together, these figures illustrate the discovery reach of nuclear interferometry, with the potential to open windows to large regions of currently uncharted phenomenological territory, thus facilitating a broader exploration of the  dark matter landscape. As a result of their differing operating conditions,  the ionic and atomic nuclear interferometer realisations proposed in this work naturally target contrasting regions of parameter space and are thus complementary in their potential physics gain. Whilst the single-ion nuclear interferometer may, if operated in space, be able to achieve significantly greater peak sensitivities than an instrument deploying clouds of neutral $^{229}$Th, the experimental parameters required for this latter realisation naturally favour higher mass particles with a significantly shallower degradation of sensitivity moving in the direction of increasing ULDM mass.  

In Figs.~\ref{fig:deb} and \ref{fig:de}, we respectively show the reach of a single-ion and a neutral atom nuclear interferometer to ULDM coupling to photons, whose future experimental landscape is expected to be most strongly constrained by long-baseline $^{87}$Sr atomic clock interferometers and nuclear-atomic optical clock comparisons. Due to the intrinsically limited ion flux, we find that terrestrial single-ion nuclear interferometers are unable to fully benefit from the increased sensitivity of the nuclear clock transition to the variation of fundamental constants, and are  likely to be out-performed by conventional $^{87}$Sr-based terrestrial interferometers, provided that the target atomic fluxes (as listed in Tab.~\ref{aion:params}) can be reached. Much higher sensitivities could be achieved in space-based experiments with the substantial increase in baseline afforded by such realisations compensating for the high shot-noise associated with working with single ions.

In contrast, we find that the $\sim$ 4 order of magnitude  increase in sensitivity of the  ground-excited isomeric state transition to the variation of $d_e$ compared to standard atomic optical clocks  is sufficient to overcome challenges associated with the short excited state lifetime in neutral $^{229}$Th, rendering terrestrial atom-based nuclear interferometers competitive with their $^{87}$Sr clock analogues in addition to improving their reach at higher masses. Beyond that, such enhancements can be achieved with both significantly less demanding LMT requirements  and more modest baselines (as is indeed required due to the short excited isomeric lifetime), than are necessary to deploy in conventional atom interferometry. 
 
 The potential of nuclear interferometry to advance existing experimental proposals is substantially greater in searches for new physics coupling to the QCD sector via the couplings $d_g$ and $d_{\hat{m}}$, as displayed in Figs.~\ref{fig:dgb}, \ref{fig:dmb}, \ref{fig:dg} and \ref{fig:dm}.  These couplings enter the frequencies of typical atomic optical clock transitions via the ratio of the electron mass to the nucleus mass ($m_e/m_{\textnormal{nuc}}$) in the reduced mass of the system, rendering their contribution suppressed by approximately 5 orders of magnitude with respect to that of $d_e$. As such, the sensitivity of conventional atomic clock interferometers to such effects is not appreciable and for consistency with literature is not shown in these figures. We note that in Ref.~\cite{Banerjee:2023bjc} it was shown that in heavy-nucleus optical clocks such as  $^{171}$Yb$^{+}$, the transition frequency becomes sensitive to the nuclear-charge radius, leading to an enhanced  sensitivity to the couplings $d_g$ and $d_{\hat{m}}$ by $\sim 2$ orders of magnitude. Such systems could in principle be deployed in clock atom interferometry experiments as recently suggested in Ref.~\cite{Zhao:2024tvg} where oscillations of the nuclear charge radius in a space-based atom interferometry experiment using a (neutral) Yb source were considered. We emphasise that the intrinsic sensitivity of this transition to variations of $d_g$ and $d_{\hat{m}}$ is still many orders of magnitude below that of the nuclear clock however. 
 
We highlight that the deployment of the nuclear clock in an interferometry-based search for ULDM as proposed in this work is complementary to its usage in nuclear-atomic optical clock frequency comparison experiments. Indeed, as can be seen in the plots, the nuclear interferometer pushes the potential parameter space coverage closer to, and in the case of $d_{\hat{m}}$ into, more theoretically motivated regions where the couplings can be deemed technically natural. The estimated extension in reach of the nuclear interferometer over nuclear-atomic clock comparisons is, however, considerably more marginal than the relative sensitivity enhancement (to the coupling to photons) that may be achieved by traditional atom interferometers compared to table-top clock frequency comparison experiments. This follows from the afore-discussed additional challenges of using $^{229}$Th (in both atomic and ionic form) for free-fall interferometry relative to e.g. a single ion clock. Given however that substantial R\&D is still required to establish the nuclear clock as a competitive frequency standard, an exploration of all potential ways to probe ULDM-induced variations of fundamental constants with $^{229}$Th is warranted at this stage. Indeed, the precise requirements on the nuclear transition (e.g. frequency stability and uncertainty etc.) are likely to be different for  a nuclear interferometer (particularly in the atomic case) and a nuclear-atomic clock comparison experiment, since their sensitivity is ultimately driven by very different considerations. Given the different challenges to be overcome, it is not clear at present which experiment may be able to reach its forecasted sensitivity first. One further opportunity to probe varying fundamental constants with $^{229}$Th, albeit at higher frequencies, could be with a M\"{o}ssbauer inspired set-up similar to that proposed in Ref.~\cite{Banerjee:2024bkp}. 

The entirety of the discussion up to this point has been based on operating the interferometer in `broadband' mode. In principle, one could alternatively operate in  `resonant' mode~\cite{Graham:2016plp} in which the interferometer space-time diagram comprises $Q$ closed diamonds each of duration $2T$. This results in a $Q$-factor sensitivity enhancement to the mass $m_{\phi}$ = $\pi/T$ over a mass range $\pi/(QT)$ relative to a broadband search with the same experimental parameters.  Such schemes are, in the context of conventional atom interferometers, thought to be of particular benefit to space-based designs in which the atom interrogation is confined to the interior of the satellites, greatly restricting the LMT order $n$ that can be employed \cite{Graham:2017pmn,Graham:2016plp}. With the on-resonance sensitivity scaling linearly with both $Q$ and $n$ in resonant mode (see Ref.\ \cite{Badurina:2021lwr}), 
it is possible to exploit the parameter $Q$ whilst maintaining a sufficiently low $n$ so as to adhere to the geometric constraints in order to reach considerably greater sensitivities than could be achieved with broadband operation. In Ref.\ \cite{Arvanitaki:2016fyj}, it was shown that systematically scanning through a band of resonant masses over the course of the experimental campaign by altering $Q$ and $T$ could extend the reach of space-based $^{87}$Sr interferometers to masses in this band relative to the equivalent broadband search. Given that resonant mode operation involves an increased number of atom-laser interactions, for the atom-based nuclear interferometer the low $\pi$-pulse efficiencies mean such schemes are unlikely to be of any benefit. In contrast, for ion-based interferometers we do expect a scanned resonant search to yield increased sensitivities, particularly to masses at the higher end of the band. Due to the dependence of the shot-noise on the total duration of the interferometer sequence, which in resonant mode is increased by a factor $Q$ to $2QT$, the relative enhancement over the analogous broadband search will not be as significant as that which may be achieved for $^{87}$Sr atom interferometers however. 

For the purpose of illustrating the experiment, we have focused on just one specific model of ULDM. Our treatment here was not intended to be exhaustive, but to showcase the potential of the experiment and pave the way towards more complete phenomenological studies in the future. Indeed, the nuclear interferometer could be used entirely analogously to investigate any other new physics model which induces periodic variations in the thorium transition energy. Both QCD axion DM and ALP DM which couple to the SM via pseudo-scalar interactions have been shown to generate such oscillations at quadratic order in the axion field \cite{Kim:2022ype,Beadle:2023flm,Kim:2023pvt}, thus making for alternative theoretical targets. The signal imprinted in a nuclear interferometer by a quadratically coupled field takes a slightly modified form compared to the linear case: to exemplify this, we make projections for (specifically)  QCD axion DM in Appendix~\ref{ap:axion}, exploiting the quadratic coupling to hadrons which emerges below the QCD scale as a consequence of the defining  axion-gluon coupling.

Whilst the reach of the proposals presented here remains limited by the high shot-noise and the short lifetime of the excited state respectively, this work demonstrates that there is nonetheless the potential, provided experimental challenges can be overcome in the future, to access new, previously uncharted, phenomenological territory with the nuclear interferometer. Not only does this serve as testament to the unique and exciting 
role that $^{229}$Th may play in our future explorations of new physics, but it motivates a closer investigation of its potential  within the field of long-baseline interferometry. For example, in the case of single-ions, one might explore different interferometer geometries to enable the use of greater ion fluxes, and in neutral atoms, alternative interferometry schemes deploying floquet atom optics \cite{Wilkason:2022yej} could be considered as a means to achieve higher $\pi$-pulse efficiencies.  

A further interesting line of enquiry could be to investigate the possibility of binding $^{229}$Th into a molecule. Not only would it be necessary to verify that the nuclear transition, and its enhanced sensitivity to the variation of fundamental constants remains intact in such a system, but that both the dissociation and ionisation energies of the molecules lie above the nuclear transition energy. To fully harness the benefits that working with a neutral species presents in terms of the shot-noise,  it would be desirable for the excited nuclear state to be long lived, avoiding the complications associated with spontaneous decay seen in the atomic case. 
\section{Conclusions}
\label{sec:sum}
With an extraordinarily broad landscape of viable candidates which give rise to a diverse range of phenomenological signatures,  new ideas for the detection of DM are required in order to maximise experimental coverage of the vast available parameter space. In this work we propose two possible realisations of a single-photon interferometry experiment based on  the nuclear clock transition in $^{229}$Th as a  probe of ULDM with scalar couplings to the SM, making use of  single-ions and clouds of neutral atoms respectively. Although working with either source type presents a unique set of experimental challenges, we find that the enhanced sensitivity of the nuclear clock transition to the variation of fundamental constants offers the means to access large regions of uncharted parameter space  over a wide range of frequencies, providing experimental challenges can be met in the future. 

 In particular, nuclear interferometry could offer a unique opportunity to probe scalar couplings of ULDM to gluons and quarks, with a sensitivity exceeding existing and proposed experiments over a range of frequencies.  Employing the nuclear clock in this way complements its usage in nuclear-atomic optical clock frequency comparisons, with the peak sensitivity moving towards more theoretically motivated regions of parameter space. Although by no-means fully optimised, this proposal highlights the important role which $^{229}$Th could play in future searches for new physics, motivating a broader investigation into its potential capabilities both within interferometry and beyond.

\acknowledgments{
This work is supported by the CERN Quantum Technology Initiative. We are incredibly grateful to Leondardo Badurina and Thomas Hird for discussions regarding the challenges of using sources with short excited state lifetime in atom interferometry and to Sebastian Ellis for raising the issue of magnetic field noise after submission of V1, reflected in the extended discussion in V2. We also thank Dennis Schlippert, Marianna Safronova, Masha Baryakhtar, Gilad Perez and Ekkehard Peik for invaluable feedback on early versions of this manuscript, Michael Doser for discussions on the nuclear clock and John Carlton for insight into the AION experiment and gravitational gradient noise. We thank Jeremiah Mitchell for discussions regarding the treatment of noise sources and  Nathaniel Sherrill and Abhishek Banerjee for details on the analyses of atomic clock comparison experiments. HB acknowledges partial support from the STFC HEP Theory Consolidated grants ST/T000694/1 and ST/X000664/1 and thanks other members of the Cambridge Pheno Working Group for useful discussions. EF acknowledges funding by the Deutsche Forschungsgemeinschaft (DFG, German Research Foundation) under Germany’s Excellence Strategy – EXC-2123 QuantumFrontiers – 390837967, and via the SFB/CRC 1227 (DQ-mat) -- Project-ID 274200144 -- of the DFG.
This work has been partially funded by the Deutsche Forschungsgemeinschaft
(DFG, German Research Foundation) - 491245950.
}


\appendix
\section{Sensitivity to ULDM Couplings}\label{ap:uldmsens}
Here we briefly outline how the experimental sensitivity given in Eq.~\ref{eq:sens} follows from  Eq.~\ref{eq:amp}, the signal amplitude for a single measurement. The experimental sensitivity takes into account a total of $N$ phase measurements made continuously at a regular sampling rate $1/\Delta t$ over the total experiment run time $T_{\textnormal{int}}$. We denote this series of measurements  as $\{\Phi_m\}$, where $m$ labels the phase recorded from the interferometer sequence starting at time $m \Delta t$. 

Such a discrete set of sequential measurements can be conveniently characterised by its power spectral density (PSD), which contains information on both the amplitude and the frequency spread of the data. It is formally constructed from the discrete Fourier transform of the sequence of measurements which itself can be calculated according to 
\begin{equation}
\tilde{\Phi}^k  = \sum_{m=0}^{N-1} \Phi_m \textnormal{exp}\left(-\frac{2\pi i m k }{N}\right)~.\end{equation} 
Here $k$ takes integer values between 0 and $N-1$ and labels the discrete (angular) frequencies $\omega = 2\pi k / T_{\textnormal{int}}$ with resolution $\Delta \omega = 2 \pi/T_{\rm int}$.

Using this definition, the signal PSD, $S_s^k$, is  
\begin{equation}
S^k_s = \frac{(\Delta t)^2}{T_{\textnormal{int}}} \left | \tilde{\Phi}^k \right |^2~.
\end{equation}

In the limit that $T_{\textnormal{int}} < \tau_c$, the ULDM signal is contained within a single frequency bin. In this instance the signal PSD takes the form
\begin{equation}
S_s(\omega) \sim  \begin{cases}
			T_\textnormal{int}| \overline{\Phi}_s |^2, & \omega = \omega_0 \\
            0, & \omega \neq \omega_0
		 \end{cases} ~, 
\end{equation}
where for convenience we have rewritten the PSD as a (continuous) function of angular frequency, $\omega = 2\pi k / T_{\textnormal{int}}$.   

 The SNR can then be constructed  as the ratio of the signal and noise PSDs evaluated at $\omega_0$ 
\begin{equation}
\label{eq:SNR}
\textnormal{SNR} = \frac{S_{\textnormal{s}} (\omega_0)}{S_\textnormal{n}(\omega_0)}~,
\end{equation}
where, assuming the experiment is shot-noise dominated, the noise power spectrum  $S_{\textnormal{n}}$ is flat, and given by Eq.~\ref{eq:noise} and Eq.~\ref{eq:SN} for the ionic and atomic realisations respectively. 

The  sensitivity to the couplings of interest in this regime is thus
\begin{equation}
d_{\phi} \simeq \frac{\sqrt{\textnormal{SNR}}}{\overline{\Phi}_R} \times \sqrt{ \frac{S_{\textnormal{n}}}{T_{\textnormal{int}}}}~.
\label{eq:sens1}
\end{equation}

If instead $T_{\textnormal{int}} > \tau_c$, the signal PSD has a finite profile in frequency space deriving from the DM velocity distribution. In this instance the experimental sensitivity is most rigorously extracted via a likelihood analysis as in e.g. Refs.~\cite{Foster:2017hbq,Badurina:2025idj}. Alternatively one may apply Bartlett's method \cite{Badurina:2021lwr,VanderPlas_2018,Budker:2013hfa}. 


The resulting sensitivity in this case is \begin{equation}
d_{\phi} \simeq \frac{\sqrt{\textnormal{SNR}}}{\overline{\Phi}_R} \times \sqrt{ \frac{S_{\textnormal{n}}}{\sqrt{\tau_c T_{\textnormal{int}}}}}~.
\label{eq:sens2}
\end{equation} 
Combining this result with Eq.~\ref{eq:sens1} leads directly to Eq.~\ref{eq:sens}.

\section{Derivation of Eq.~\ref{eq:split}}
\label{app:deri}
In this Appendix we derive Eq.~\ref{eq:split}, the energy splitting between the desired and competing processes for the first selective $\pi$-pulse of the interferometry sequence (the second pulse in Fig.~\ref{fig:spacetime}). This pulse is fired in the $-z$ direction and is intended to induce stimulated emission in the upper arms of the interferometers whilst not exciting the lower arms. 

Suppose that at the time this $\pi$-pulse is fired, the atom cloud is moving  upwards with speed $v_0$, corresponding to a momentum $p_0 = m_{\rm Th} v_0$. At this point
the atom is in a  coherent superposition of ground and excited wavepackets, created by the preceding $\pi/2$-pulse. Conservation of momentum during that process means that the excited component (corresponding to the upper arm of the interferometer) has acquired a momentum $k$ in the $+z$ direction relative to the ground-state component (lower arm).

Labelling states by both their internal state (either `e' for excited or `g' for ground), and momentum (in the $+z$-direction), we can then write the initial state for the desired interaction of the $\pi$-pulse with the upper arm as $\ket{e, p_0 + k}$. The final state following the intended stimulated emission process is $\ket{g, p_0 + 2k}$. Being an emission process, the requisite (resonant) photon energy to induce this process is
\begin{align}
E_1^{\rm desired} &= E(\ket{e, p_0 + k}) - E(\ket{g, p_0 + 2k}) \\  &=
\omega_N - \frac{p_0 k}{ m_{\rm Th}} - \frac{3k^2}{2 m_{\rm Th}}~.
\end{align}

Now consider the competing absorption process in  the lower arm. In this case the initial state is $\ket{g, p_0}$. Since the laser pulse propagates in the $-z$ direction, the atom recoil is also in this direction. The final state is therefore $\ket{e, p_0 - k}$. As such
\begin{align}
E_1^{\rm competing} &= E(\ket{e, p_0 - k}) - E(\ket{g, p_0}) \\  &=
\omega_N -  \frac{p_0 k}{ m_{\rm Th}} + \frac{k^2}{2 m_{\rm Th}}~.
\end{align}
It then follows that 
\begin{equation}
\Delta E_{1} = \left | E_{1}^{\rm competing} -  E_{1}^{\rm desired} \right|  = \frac{2   k^2}{ m_{\rm Th}}~.
\end{equation}

\section{$\pi$-pulse efficiencies}\label{ap:pi}
The behaviour of a 2-level atomic or nuclear system coherently interacting with an external electromagnetic field (i.e. a laser) over time can be described by the optical Bloch equations (see e.g. Refs.~\cite{steck2007quantum, Scully_Zubairy_1997}). These equations model the time evolution of a density operator $\hat{\rho}$ which encodes the occupation probabilities of the ground and excited states. When the field is on resonance with the transition, in the absence of spontaneous decay, the population probabilities of the ground and excited state oscillate at the so-called Rabi frequency, $\Omega$, such that after a time $t_{\pi} = \pi / \Omega$, a system initialised in the ground state has a 100$\%$ probability of having transitioned to the excited state. Pulses of this duration are commonly referred to as $\pi$-pulses, and as discussed in the main text, play a crucial role in the coherent transfer of populations in matter-wave interferometry. 

To estimate the $\pi$-pulse efficiency in the presence of spontaneous decay, one can explicitly introduce spontaneous decay to the optical Bloch equations and solve for the probability of being in the excited state as a function of time assuming that the system is initialised in the ground state and that the electromagnetic field (i.e. the laser) is resonant with the transition. This calculation, which can be performed analytically, is sometimes known as Torrey's solution \cite{PhysRev.76.1059} and is derived in full in Refs. \cite{steck2007quantum,NOH20102353}. For a comprehensive discussion of these equations in the context of nuclear transitions, we refer the reader to Ref.~\cite{vonderWense:2020kck}. 

Evaluating the excited state probability at $t_\pi = \pi / \Omega$ yields 
\begin{equation}
\begin{split}
&P \hspace{0.5em}\Big|_{t = t_{\pi}} = \frac{\Omega^2}{2 \Omega^2 + \Gamma_a^2} \hspace{0.5em}\times \\ &\left[1 - e^{-\frac{3\Gamma_a \pi}{4 \Omega}}\left[\cos{\left(\frac{\Omega_{\Gamma}}{\Omega}\pi\right) }+ \frac{3\Gamma_a}{4\Omega_\Gamma}\sin{\left(\frac{\Omega_{\Gamma}}{\Omega}\pi\right)}\right]\right]~,
\end{split}
\label{eq:piprob}
\end{equation}
where 
\begin{equation}
\Omega_\Gamma = \Omega \sqrt{1 - \left(\frac{\Gamma_a}{4 \Omega} \right)^2}~.
\end{equation}

In quoting this result we have implicitly taken the limit that the bandwidth of the exciting laser, $\Gamma_l$, is significantly less than the natural width of the transition  $\Gamma_a \sim 16$ kHz, such that the rate of decoherence in the optical Bloch equations:  $\tilde{\Gamma} = (\Gamma_a + \Gamma_l$)/2 is entirely due to spontaneous decay i.e. $\tilde{\Gamma} \sim \Gamma_a/2$. It is this regime which maximises the $\pi$-pulse efficiency.

We take  Eq.~\ref{eq:piprob} - i.e. the probability that a pulse of duration $t_\pi$ results in a population inversion - to be the $\pi$-pulse efficiency that we use in making our sensitivity forecasts and emphasise that it is entirely governed by the ratio $\Omega / \Gamma_a$.

\section{AION Prospective Sensitivity} \label{ap:aion}

The sensitivity curves labelled TVLBAI-10, TVLBAI-100 and TVLBAI-km  displayed in Fig.~\ref{fig:de} refer specifically to   terrestrial atom interferometers operating on the 698 nm 5s$^2$ $^1S_0 \leftrightarrow$ 5s5p $^3P_0$ clock transition in $^{87}$Sr assuming the experimental parameters detailed in Tab.~\ref{aion:params}. These choices match those that may be implemented in future vertical atom gradiometers such as AION. In particular, the TVLBAI-km parameters correspond to those used in the plots presented by the terrestrial very-long-baseline atom interferometry community in Ref.~\cite{Proceedings:2023mkp}. The TVLBAI-10 and TVLBAI-100 parameter sets match configurations considered by the AION collaboration in Refs.~\cite{Badurina:2021lwr} and \cite{Badurina:2019hst} respectively. The parameters chosen for the SB-VLBAI configuration are appropriate for a broadband search at AEDGE with the interrogation region situated exterior to the satellites. In this respect the values of $n$ and $T$
are selected to achieve a maximum cloud separation of $\sim$ 100\,m whilst paying heed to the constraints on the maximum number of laser pulses (1000) and the maximum interferometer sequence duration (300\,s) that could be implemented at AEDGE as stated in Ref.~\cite{AEDGE:2019nxb}. For each of these forecasts, a total campaign duration of 10$^{8}$\,s is assumed.   
\begin{table*}
\begin{center}
\begin{tabular}{ |c|c|c|c|c|c| } 
 \hline
   Setup & $L$ (m)  & $T$ (s) &  $n$& $\sqrt{S_{\textnormal{n}}}$ (Hz$^{-1/2}$) & $\Delta z $ (m) \\ 
   \hline
TVLBAI-10 & 10  &  0.3 & 1000& 10$^{-4}$ & 7.8 \\
  \hline
TVLBAI-100 &  100  & 1.4 &  1000 &10$^{-4}$& 90 \\
  \hline
TVLBAI-km &  1000  & 1.7 & 2500& $10^{-5}$& 970 \\
 \hline
 SB-VLBAI & 4.4 $\times 10^7$ & 100 & 150 & $10^{-5}$& 4.4$\times 10^7$ \\
 \hline
\end{tabular}
\caption{The parameters used in the $^{87}$Sr source terrestrial and space-based very-long-baseline atom interferometer sensitivity curves displayed in Fig.~\ref{fig:deb} and Fig.~\ref{fig:de}.  These values correspond to configurations considered by proposed long-baseline atom interferometry experiments as detailed in the text.   }
\label{aion:params}
\end{center}
\end{table*}

The projections are calculated using Eq.~\ref{eq:amp} but with  $\overline{\Delta\omega}_N$ replaced by 
\begin{equation}
    \overline{\Delta\omega}_A = \omega_A \kappa \phi_0 \left( 2 + \xi_A \right)d_e~,
\end{equation}
where $\omega_A$ = 1.78\,eV \cite{Hachisu:17} and $\xi_A \sim 0.06$ \cite{Angstmann_2004}. 

\section{ Sensitivity to the (light) QCD axion}
\label{ap:axion}

 For the purpose of illustrating the concept of nuclear interferometry we have thus far focused on the sensitivity to ULDM with linear scalar couplings to the SM as defined by the  Lagrangian density in Eq.~\ref{eq:lagrangian}. In this appendix, we demonstrate that the proposed experiment is not exclusive to this particular model and can be used to probe other new physics scenarios which induce the apparent oscillation of fundamental constants. One particularly straightforward example of such a model (which we will not consider further here) is that of spin-0 scalar ULDM with quadratic couplings to the SM see e.g. Refs.~\cite{Jiang:2024agx,Banerjee:2022sqg}.  
 
 A further instance is that of QCD axion DM, which at phenomenologically viable masses, also behaves as a persistent classic wave, albeit with \textit{pseudo}-scalar couplings to the SM. It was recently pointed out that  QCD axion DM generates periodic temporal variations of atomic transition energies at quadratic order in the axion field by inducing oscillations in hadronic masses and other nuclear quantities\footnote{In this work we focus on the oscillations in the $^{229}$Th transition energy that result from the quadratic couplings  of the axion to hadrons that emerge below the QCD scale from the defining axion-gluon coupling.  Oscillations can also be induced by the quadratic coupling of the QCD axion to photons which arises at 1-loop level due to threshold corrections \cite{Beadle:2023flm,Kim:2023pvt}.  These are loop-suppressed relative to the hadronic couplings and we thus do not consider these further here. } \cite{Kim:2022ype}.  
 
The defining feature of QCD axion models is an anomalous coupling between the \textit{pseudo}-scalar axion field, $a$, and the SM gluon fields  \begin{equation} \label{eq:def_ax}
 \mathcal{L} \in \frac{g_s^2}{32 \pi^2}\frac{a}{f_a} G^{\mu \nu}\widetilde{G}_{\mu \nu}~,\end{equation} where $f_a$ is the axion decay constant, $g_s$ the strong coupling constant and $G$, ($\widetilde{G}$) the (dual) gluon field strength tensor. Below the QCD confinement scale, this interaction induces a quadratic coupling of the axion to hadronic states. In particular, the pion mass obtains a dependence on the axion field according to  
 \begin{equation}
 m_\pi^2 (\theta_{\textnormal{eff}}) = \frac{\Lambda^3_{\textnormal{QCD}}}{f_\pi^2}\sqrt{m_u^2 + m_d^2 + 2 m_u m_d \cos (\theta_{\textnormal{eff}})}~,\end{equation}
 where, for a QCD DM axion with mass $m_a$, \begin{equation}\label{eq:thetaeff}
 \theta_{\textnormal{eff}} = \frac{(a - \langle a \rangle)}{f_a} = \frac{\sqrt{2\rho_{DM}}}{m_a f_a} \cos (m_a t)~.
\end{equation}
Given that it behaves as a coherently oscillating classical wave, QCD axion DM  thus induces an oscillatory contribution to the pion mass at quadratic order in the axion field: 
\begin{equation}
\label{eq:masspi}
\frac{\Delta m_\pi^2}{m_\pi^2} \sim \frac{m_u m_d \theta_{\textnormal{eff}}^2(t)}{2(m_u + m_d)^2}~.
\end{equation}
This in turn causes the $^{229}$Th nuclear transition energy to oscillate as  \cite{Kim:2022ype} \begin{equation}
    \frac{\Delta \omega_N}{\omega_N} \sim (2 \times 10^5) \frac{\Delta m_\pi^2}{m_\pi^2}~.
\end{equation}

Combining Eq. \ref{eq:thetaeff} and Eq. \ref{eq:masspi}, one can read off the amplitude of
$\Delta \omega_N$ (defined in analogy to Eq.~\ref{eq:ampw}) to be
\begin{equation}
\overline{\Delta \omega_N} = \omega_N (2 \times 10^5)\left(\frac{m_u m_d}{(m_u + m_d)^2}\right)\left(\frac{\rho_{DM}}{m_a^2 f_a^2}\right)~.
\end{equation}

When the transition energy depends quadratically on the oscillating field as we have here, the single-shot signal amplitude (c.f. Eq.~\ref{eq:amp}) for a gradiometer set-up takes the modified form
\begin{equation}
\begin{split}
\overline{\Phi}_{s,a} = 2 \frac{\overline{\Delta \omega_N}}{m_a} \frac{\Delta z}{L} \sin \left[ m_a (T - (n-1) L)\right]  \\ \times  \sin \left[ m_a n L \right]  \sin \left[ m_a T \right]~.
\end{split}
\label{eq:amp2}
\end{equation}

Defining, in analogy to Eq.~\ref{eq:fac}, 
\begin{equation}
    \overline{\Phi}_{s,a} = \frac{\overline{\Phi}_{R,a}}{f_a^2}~,
\end{equation}  it follows that the minimum value of the inverse  axion decay constant that could be detected at a given SNR following a measurement campaign of duration $T_{\textnormal{int}}$ is

\begin{equation}
\frac{1}{f_a^*} \simeq \sqrt{\frac{\sqrt{\textnormal{SNR}}}{\overline{\Phi}_{R,a}} \times \sqrt{ \frac{S_{\textnormal{n}}}{T_{\textnormal{int}}}\times \textnormal{max}\left(1,\sqrt{\frac{T_{\textnormal{int}}}{\tau_c}}\right)}}~.
\end{equation}
where, as before, $S_n$ denotes the flat shot noise PSD as given in Eq.~\ref{eq:noise}. 

\begin{figure}
\centering
\includegraphics[width=0.45\textwidth]{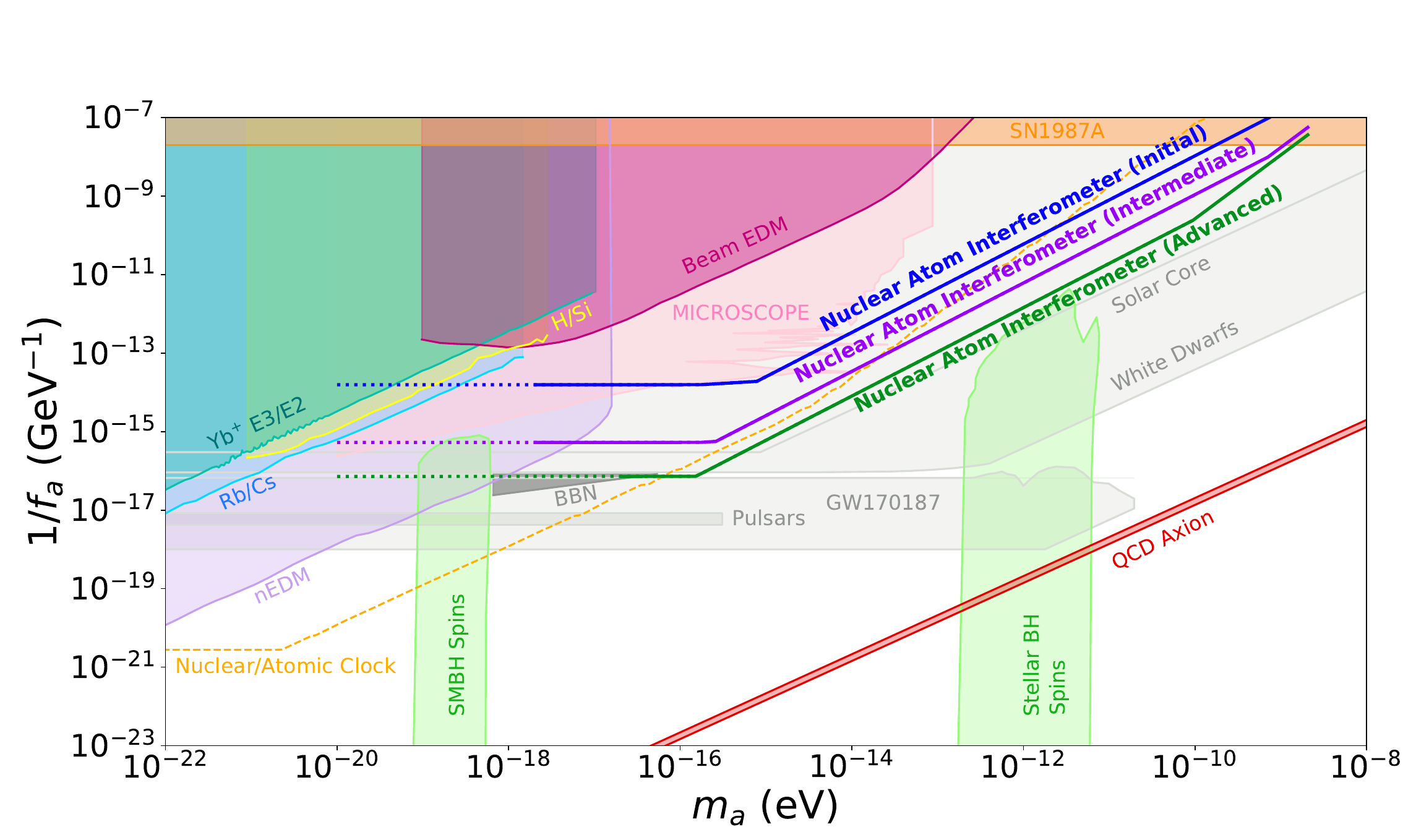}
\caption{Shot-noise limited projected sensitivity of the  nuclear interferometer configurations deploying neutral atoms detailed in Tab.~\ref{params_atom} to QCD axion DM via its quadratic couplings to hadrons. The curves show the minimum value of $1/f_a$  that could be detected at a SNR = 1 following an experimental campaign of duration $T_{\textnormal{int}}=10^8$\,s and assuming that photo-ionisation has been successfully suppressed. The dotted portion of the curves indicates where the sensitivity may be adversely impacted by GGN. Also shown are existing bounds from astrophysics and other experiments as detailed in the main text, in addition to the forecast for nuclear-atomic clock frequency comparisons. The red QCD axion band indicates the expected parameter space coverage of standard QCD axion models in which the axion mass is generated by the anomalous coupling to quarks.}
\label{fig:ax_atoms}
\end{figure}

\begin{figure}
\centering
\includegraphics[width=0.45\textwidth]{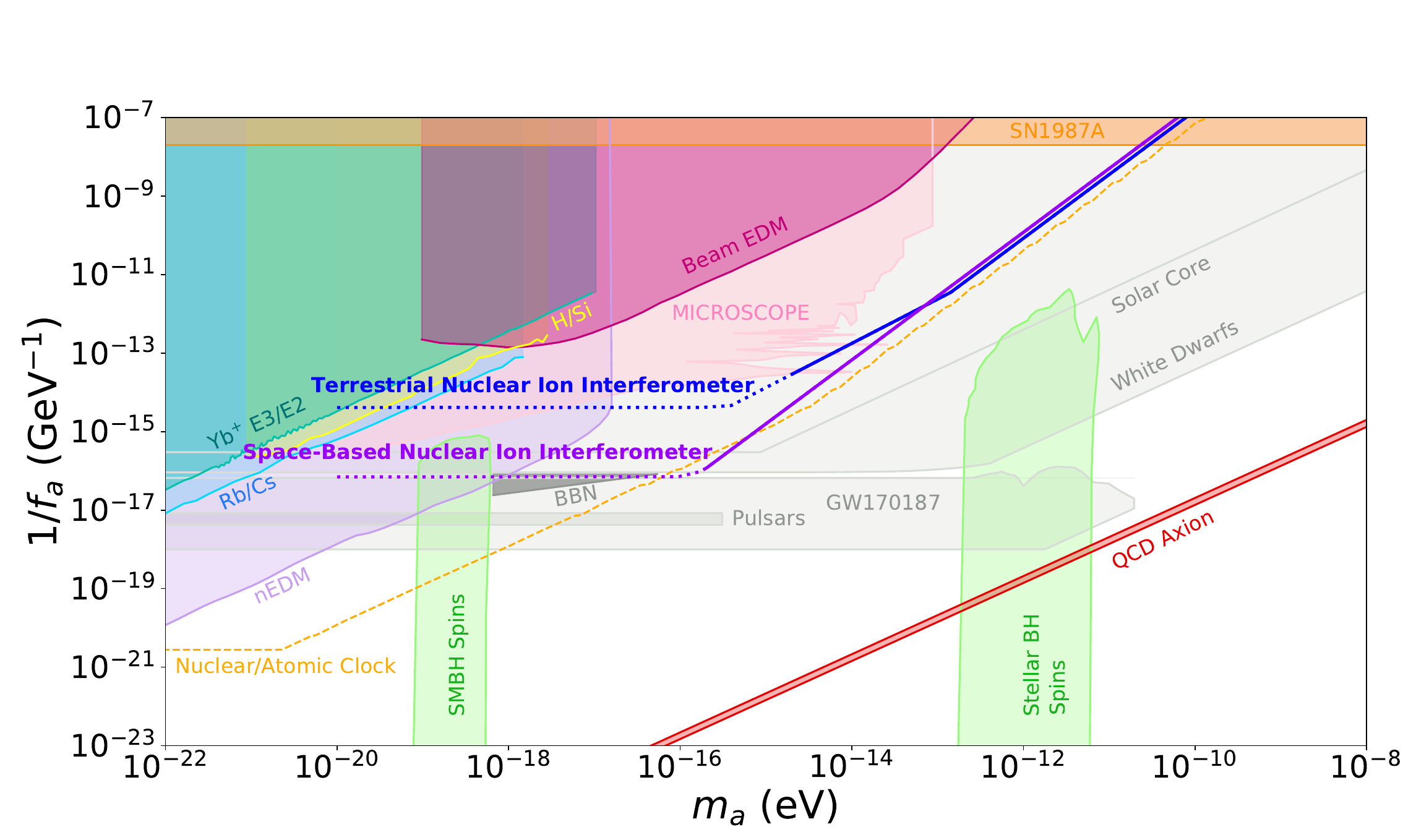}
\caption{Shot-noise limited projected sensitivity of the single-ion nuclear interferometer configurations detailed in Tab.~\ref{paramsion} to QCD axion DM via its quadratic couplings to hadrons. The curves show the minimum value of $1/f_a$  that could be detected at a SNR = 1 following an experimental campaign of duration $T_{\textnormal{int}}=10^8$s. The dotted portion of the curve indicates where the sensitivity may be adversely impacted by magnetic field noise. Also shown are existing bounds from astrophysics and other experiments as detailed in the main text, in addition to the forecast for nuclear-atomic clock frequency comparisons. The red QCD axion band indicates the expected parameter space coverage of standard QCD axion models in which the axion mass is generated by the anomalous coupling to quarks.}
\label{fig:ax_ions}
\end{figure}

Figures~\ref{fig:ax_atoms} and \ref{fig:ax_ions} show the projected sensitivity of the atomic and ionic nuclear interferometer configurations (as detailed in the main text)  to  QCD axion DM in the ($m_a$,1/$f_a$) plane\footnote{Such plots can be challenging to interpret from a theoretical standpoint given the additional model building required to reach parameter space outside the QCD axion band, and should therefore be viewed purely as an comparison of experimental sensitivities.}, at an  SNR of 1 and assuming a total integration time of 10$^8$ s. The noise spectrum $S_n$ for each instrument is assumed to be shot-noise dominated and is modelled in the same way as for the case of linear couplings as described in the main text (i.e. Eq.~\ref{eq:noise} and Eq.~\ref{eq:SN} for the ionic and atomic realisations respectively). Dotted lines are once again used to indicate where, without the application of additional practices to mitigate GGN and magnetic noise in the case of the atom and ion realisations respectively, the experiment may depart from the shot-noise limited regime. In computing the mass at which this may occur it is important to remember that since the coupling is  quadratic, the signal frequency corresponds to twice that of the axion mass. 

For comparison we show the limits from an oscillating neutron EDM ~\cite{Abel:2017rtm} and beam based EDM experiment~\cite{Schulthess:2022pbp}, in addition to a Rb/Cs hyperfine frequency clock comparison~\cite{Hees:2016gop} and a frequency comparison of a hydrogen maser and a silicon cavity  (H/Si) \cite{Kennedy:2020bac}, using the limits derived in \cite{Kim:2022ype}. The limits from a frequency comparison of the E3/E2 transitions in Yb$^{+}$~\cite{Banerjee:2023bjc} and MICROSCOPE~\cite{Gue:2025nxq} are also displayed. In addition to laboratory experiments, we also show  constraints from Supernova 1987~\cite{Springmann:2024ret}, white dwarfs~\cite{Balkin:2022qer}, the binary neutron star gravitational wave event GW170817~\cite{Zhang:2021mks} and big bang nucleosynthesis (BBN) \cite{Blum:2014vsa}. Also displayed are bounds from binary pulsars~\cite{Hook:2017psm} and a neutrino line from the solar core~\cite{Hook:2017psm}, using the corrected limits presented in Ref.~\cite{DiLuzio:2021pxd} in both instances. Superradiance constraints from both stellar mass black holes and supermassive black holes (SMBH) are also shown ~\cite{Mehta:2020kwu,Baryakhtar:2020gao,Unal:2020jiy,Hoof:2024quk,Witte:2024drg}. Although there are a variety of other proposals forecast to place constraints on this parameter space  e.g. Refs.~\cite{Arvanitaki:2021wjk,Chang:2017ruk,Berlin:2022mia,JacksonKimball:2017elr},  other than the projection for the frequency comparison of a nuclear and atomic clock~\cite{Kim:2022ype}, we do not display these projections here to avoid overcrowding. 

Parameter space outside the red `QCD axion' band corresponds to models which solve the strong CP problem, but have a mass that differs (accounting for theoretical uncertainty) from that generated by the anomalous coupling to gluons (Eq.~\ref{eq:def_ax}) which defines $f_a$. Populating these regions generally requires significant model building and often suffers from naturalness problems. 

From these figures, it is evident that the proposed realisations of the nuclear interferometer presented here are only able to probe `light' QCD axions, whose masses are substantially less than that predicted from their defining QCD coupling. One example of a `natural' model in which this feat is achieved is presented in Ref.~\cite{Hook:2018jle} and subsequently developed phenomenologically in Ref.~\cite{DiLuzio:2021pxd}. 

Whilst both the atomic and ionic nuclear interferometer configurations are able to reach substantial regions of parameter space not probed by existing laboratory experiments (and are on a par with forecasts from nuclear-atomic clock comparisons), this space has already been ruled out by astrophysical considerations. A nuclear interferometer could nonetheless provide a useful independent verification of these limits in a controlled laboratory environment. 

\bibliographystyle{JHEP}
\providecommand{\href}[2]{#2}\begingroup\raggedright\endgroup


\end{document}